\pdfoutput=1

\def\be{\begin{equation}}
\def\ee{\end{equation}}
\def\bear{\begin{eqnarray}}
\def\eear{\end{eqnarray}}
\def\bdm{\begin{displaymath}}
\def\edm{\end{displaymath}}

\documentclass{osa-article}

\usepackage{comment}
\usepackage{bm}
\usepackage{hyperref}

\journal{osajournal}

\articletype{Research Article}

\begin{document}

\title{Application of Tilt Correlation Statistics to Anisoplanatic Optical Turbulence Modeling and Mitigation}


 \author{Russell C. Hardie,\authormark{1,*} Michael A. Rucci,\authormark{2} 
 \\ Santasri Bose-Pillai,\authormark{3} and Richard Van Hook\authormark{2}}

\address{\authormark{1}University of Dayton, Department of Electrical and Computer Engineering, 300 College Park, Dayton OH, USA, 45459-0232\\
\authormark{2}Air Force Research Laboratory, AFRL/RYMT, Building 620, 2241 Avionics Circle, Wright-Patterson AFB, Ohio, USA, 45433\\ 
\authormark{3}Air Force Institute of Technology, Department of Engineering Physics, 2950 Hobson Way, Dayton, Ohio, USA 45433}

\email{\authormark{*}Corresponding author: rhardie1@udayton.edu} 



\begin{abstract}
Atmospheric optical turbulence can be a significant source of image degradation, particularly in long range imaging applications.  Many turbulence mitigation algorithms rely on an optical transfer function (OTF) model that includes the Fried parameter.  We present anisoplanatic tilt statistics for spherical wave propagation.  We transform these into 2D autocorrelation functions that can inform turbulence modeling and mitigation algorithms.   Using these, we construct an OTF model that accounts for image registration.  We also propose a spectral-ratio Fried parameter estimation algorithm that is robust to camera motion and requires no specialized scene content or sources.   We employ the Fried parameter estimation and OTF model for turbulence mitigation.  A numerical wave-propagation turbulence simulator is used to generate data to quantitatively validate the proposed methods.  Results with real camera data are also presented.
\end{abstract}

\noindent
© 2021 Optical Society of America. One print or electronic copy may be made for personal use only. Systematic reproduction and distribution, duplication of any material in this paper for a fee or for commercial purposes, or modifications of the content of this paper are prohibited.
\url{https://doi.org/10.1364/AO.418458}



\section{Introduction}
\label{intro_section}  

In long-range imaging, 
atmospheric optical turbulence can be a significant source of image degradation~\cite{ROGWEL1996}.
Therefore it is important to develop effective turbulence mitigation image restoration algorithms.
In terrestrial imaging applications, a relative wide field-of-view is used.  This leads to anisoplanatic
conditions where the acquired short-exposure images are corrupted by spatially and temporally varying warp and blur.
One simple and effective turbulence mitigation method is the 
Block Matching and Wiener Filtering (BMWF) algorithm~\cite{bmwf2017,FIFSR2019}.   
The BMWF method uses a Block Matching Algorithm (BMA) to perform dewarping on a sequence of short exposure 
frames.  The dewarped frames are then fused in a temporal average or weighted average~\cite{Karch2015}.  
Image restoration with a Wiener filter is then applied to the fused image for deblurring purposes.   

To specify the Wiener filter, a model for the optical transfer function (OTF) that relates the fused image to the undegraded truth image is required.  We use a parametric OTF model here that relies on the atmospheric coherence diameter, $r_0$, 
also referred to as the Fried parameter~\cite{FRIED1966}.   The Fried parameter governs the 
short-exposure atmospheric effects in the OTF model~\cite{FRIED1966}.   Our OTF model also takes into account how effective
the image registration is in performing atmospheric tilt correction.  For this, we define and use a parameter that we refer 
as the tilt correction factor~\cite{bmwf2017,FIFSR2019}.    
In order to use a turbulence mitigation method such as the BMWF in an automated manner, 
we need to be able to estimate the Fried parameter from the observed images
and separately determine the tilt correction factor. 
The main contribution of this paper is in the development
of automated novel ways to obtain these two parameters,
and utilize them for turbulence mitigation with the BMWF method. 

There is a great deal of interest in characterizing atmospheric optical turbulence to help understand its impact on imaging 
and optical communications~\cite{Lars2005,WangSong2018,Chatterjee2018,Santasri2018,Chatterjee2019}.   
Thus, estimating the Fried parameter is important for many other applications in addition to the BMWF algorithm.  
This parameter can be measured with specialized equipment such as a scintillometer that monitors intensity fluctuations,
or a differential image motion monitor~\cite{DIMM1990} that uses a point source and two apertures.
Some image-based methods have been developed that use a standard camera with specialized targets or sources~\cite{mza_delta,Gladysz2015A}.  Other methods are scene-based and use only the natural imagery acquired by an imaging sensor~\cite{Zamek2006,Gladysz2012,Gladysz2013,vonderLuhe:84,Gladysz2015B,Basu2015,McCrae2016}. Methods that explicitly address camera or 
scene motion include~\cite{Kelmelis2017,Santasri2018}.

In this paper, we present anisoplanatic tilt statistics for spherical wave propagation.  Treating the tilts as wide sense stationary (WSS), we transform the tilt correlations into 2D autocorrelation functions that can inform modeling and mitigation algorithms.  
We use the 2D autocorrelation functions here to aid in forming our OTF model that captures the level of tilt correction
provided by image registration.  We use this OTF model to develop a modified spectral ratio Fried parameter estimation algorithm~\cite{vonderLuhe:84} that utilizes the full image, is robust to camera motion, and requires no specialized scene content or sources.  
The OTF model that employs the proposed Fried parameter estimation method is applied
here for turbulence mitigation using the BMWF method.
We evaluate the efficacy of the proposed Fried parameter estimation and turbulence mitigation using both simulated and real camera datasets.  The simulated data are generated using the anisoplanatic numerical wave-propagation method 
developed by Hardie {\em et al}~\cite{HardieSimulation2017}.  These data allow for quantitative performance analysis
because ground truth images are available.

The organization of the remainder of this paper is as follows.  In section \ref{mitigation_section},
we introduce our turbulence mitigation approach as this sets the stage and provides motivation
 for the other contributions in this paper.
Anisoplanatic tilt correlation statistics are presented in Section \ref{tilt_section}.  
The application of the tilt statistics to turbulence OTF modeling 
with image registration is developed in Section \ref{otfappsec}.   
This leads to our proposed spectral-ratio Fried parameter estimation method 
in Section \ref{ratio_section}.  Experimental results are presented in Section \ref{results_section}.  
Finally, we offer conclusions in Section \ref{conclusions_section}.

\section{Turbulence Mitigation}
\label{mitigation_section}

\begin{figure}[t]
\begin{center}
{\includegraphics[trim=2cm 9cm 5cm 1cm,scale=.6]{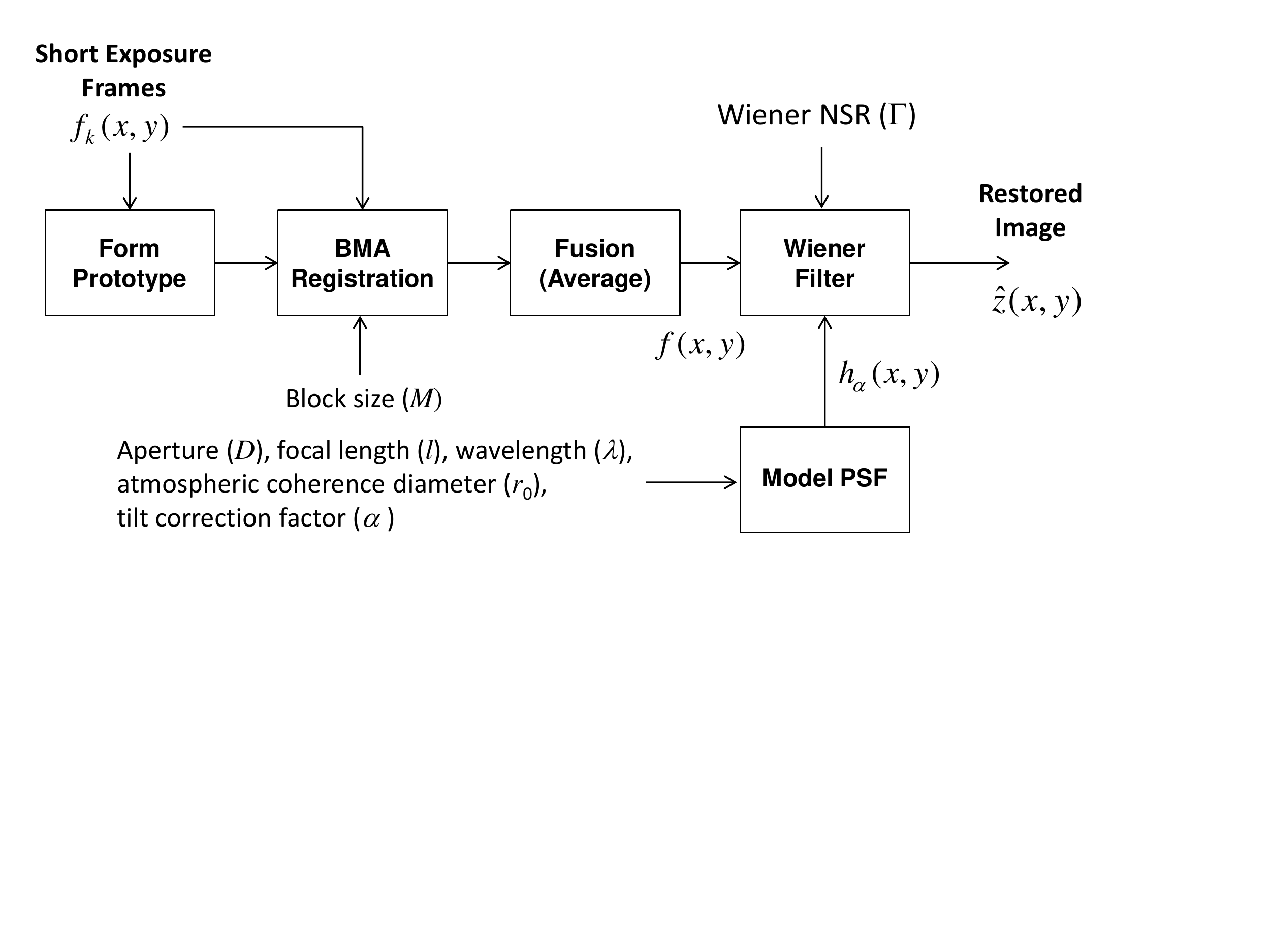}} 
\end{center}
\caption{Block matching and Wiener filtering turbulence mitigation system block diagram\cite{bmwf2017}. }
\label{Block1}
\end{figure}

A block diagram of the BMWF~\cite{bmwf2017} turbulence mitigation algorithm used here is shown in Fig. \ref{Block1}.  
The basic idea of BMWF is to use image registration to perform dewarping and then Wiener filtering to deblur.
The observed frames are denoted $f_k(x,y)$ for $k=1,2,...,K$, 
where $x,y$ are spatial coordinates and $k$ is the temporal frame index.   
The ideal image is assumed to be $z(x,y)$ and the estimate formed is denoted 
$\hat{z}(x,y)$.  

The first step illustrated in Fig. \ref{Block1} is to form a prototype image with the approximately correct geometry from the observed frames.  Because the warping shifts 
have a zero-mean, a simple average of the observed frames often produces a useful prototype\cite{bmwf2017,Fraser1999}.  This prototype is effectively a long exposure image.  It has improved geometry compared with individual frames, but has a significant amount of turbulence motion blur as well.   Global registration may be applied here prior to averaging to reduce some of the blurring and compensate for camera motion.
To further reduce turbulence motion blur, the individual observed frames are registered to the prototype using a BMA.
Let the BMA block size be defined as $(2M+1)\times(2M+1)$, for integer parameter $M$.  The BMA registered images are then fused with a simple average to form the image $f(x,y)$ as shown in Fig. \ref{Block1}.   Other fusion options are also possible~\cite{Karch2015}.
Note that $f(x,y)$ should have significantly less motion blurring than the prototype because of the BMA registration.
Ideally, the BMA registration would perfectly compensate for the turbulence warping.
However, as explained by Hardie {\em et al}~\cite{bmwf2017}, any finite-size block registration method
will provide only partial tilt correction because of the patch averaging effect~\cite{Basu2015,McCrae2016}.

Following the original BMWF development~\cite{bmwf2017}, we shall model the fused image $f(x,y)$ as 
\be
f(x,y) = h_\alpha(x,y)*z(x,y) + \eta(x,y),
\label{halphaz}
\ee
where $h_\alpha(x,y)$ is a linear shift invariant (LSI) point spread function (PSF) 
that accounts for diffraction, the average short-exposure atmospheric blurring, and
residual turbulence motion blur~\cite{bmwf2017}.  
The term $\eta(x,y)$ is assumed to be additive Gaussian noise.
In the spatial frequency domain, the model is expressed as 
\be
F(u,v) = H_\alpha(u,v)Z(u,v) + N(u,v),
\label{Halphaz}
\ee
where $u,v$ are spatial frequency variables.  The degradation OTF is given by 
\be
H_\alpha(u,v)=FT\{ h_\alpha(x,y) \},
\label{Halpha}
\ee
where $FT\left\{ \cdot \right\}$ represents the 2D Fourier transform.
The frequency spectrum of the ideal image, fused image, and noise are given by
$Z(u,v)$, $F(u,v)$, and $N(u,v)$, respectively.   What makes this model different from
traditional approaches is that it includes impact of the BMA registration on the fused image $f(x,y)$.
This unique feature is captured by the parameter $\alpha$ that
we refer to as the tilt correction factor\cite{bmwf2017}.  It provides a measure 
of how effective the BMA registration is at compensating for the warping from turbulence tilt variance.
We formally define $\alpha$ in Section \ref{tilt_section} and provide 
details on the blurring model expressed in Eqs. (\ref{halphaz}) and (\ref{Halphaz}) in Section \ref{otfappsec}.

Finally, we see in Fig. \ref{Block1} that a Wiener filter is used to deconvolve the LSI blurring from Eq. (\ref{halphaz}).
The frequency response of the Wiener\cite{Gonzalez2006} filter is given by
\be
{H_W}(u,v) = \frac{{H_\alpha{{(u,v)}^*}}}{{|H_\alpha(u,v){|^2} + \Gamma }},
\label{wiener}
\ee
where $\Gamma$ represents a constant noise-to-signal (NSR) power spectral density ratio.  
The final spatial domain BMWF estimate can then be expressed as
\be
\hat{z}(x,y)=FT^{-1}\left( H_W(u,v)  F(u,v) \right).
\label{zout}
\ee
In practice, Eq. (\ref{zout}) is computed in discrete space using the fast Fourier transform (FFT)\cite{Oppenheim2010}.

The key to the BMWF method is the PSF model used in Eq. (\ref{halphaz}), or equivalently the OTF model in Eq. (\ref{Halphaz}).
As we will see in Section \ref{otfappsec}, this model depends on optical parameters, Fried parameter $r_0$~\cite{FRIED1966}, 
and the BMA tilt correction factor $\alpha$ \cite{bmwf2017}.  
The optical parameters are generally known for a given imaging system.  Thus, the challenge in modeling the blurring PSF/OTF is in estimating $\alpha$ and $r_0$.   The main contribution of this paper is the presentation and analysis of new methods to estimate these two parameters from a sequence of short exposure images.  We will present a method in Section \ref{residualsec} that computes $\alpha$ only from knowledge of key optical parameters and the block size parameter $M$.  Next, we present a modified spectral ratio method to estimate $r_0$  in Section \ref{ratio_section}.  This method uses $\alpha$ and a sequence of short exposure images.  
One of the novel features of the proposed $r_0$ estimation method relates to how we address
camera motion and image registration using $\alpha$. 
Furthermore, it utilizes the full image size and does not require any specialized sources or target images.  Our approaches for computing $\alpha$ and estimating $r_0$ rely on anisoplanatic tilt correlation statistics.  Thus, we 
turn our attention to these statistics in the following section.

\section{Anisoplanatic Tilt Correlation Statistics}
\label{tilt_section}

\subsection{Parallel and Perpendicular Separation Tilt Correlations}

Let the two-axis Zernike-tilt (Z-tilt) angle vector be expressed as
${\tilde{\bf z}}  ( {\bm \theta}  )    =  [{{\tilde{z}_x}({{\bm{\theta }}})}  ,  {{\tilde{z}_y}({{\bm{\theta }}})}  ]^T$ 
 for a point source originating from an angle of ${\bm \theta}=[\theta_x,\theta_y]^T$. 
For a spherical wave characterized by the Kolmogorov power spectrum, the tilt correlation has
 been derived by Basu (now Bose-Pillai) {\em et al}\cite{Basu2015,Santasri2018} 
 using methods based on those by Fried\cite{Fried1976} and Winick\cite{Winick88}. 
  The result for the total tilt correlation for two point sources originating from
 angles $\bm{\theta}_1$ and $\bm{\theta}_2$ is given by 
 \be
 \label{totalcorr}
 \begin{array}{l}
   {{\tilde r}_T}(\theta ) = \left\langle { {{\tilde{\bf z}} }({{\bm{\theta }}_1}) \cdot {{\tilde{\bf z}}}({{\bm{\theta }}_2})} \right\rangle  = \left\langle {{\tilde{z}_x}({{\bm{\theta }}_1}){\tilde{z}_x}({{\bm{\theta }}_2}) + {\tilde{z}_y}({{\bm{\theta }}_1}){\tilde{z}_y}({{\bm{\theta }}_2})} \right\rangle  =  \\ 
 \left( { - \frac{{2.91}}{2}} \right){\left( {\frac{{16}}{\pi }} \right)^2}{D^{ - 1/3}}\int\limits_{z = 0}^L {C_n^2} (z)\int\limits_{\vartheta  = 0}^{2\pi } {\int\limits_{u = 0}^1 {\left[ {u{{\cos }^{ - 1}}(u) - {u^2}\left( {3 - 2{u^2}} \right)\sqrt {1 - {u^2}} } \right]} }  \times  \\ 
 {\left[ {{u^2}{{\left( {\frac{z}{L}} \right)}^2} + {{\left( {\frac{{L - z}}{D}\theta } \right)}^2} + 2u\left( {\frac{z}{L}} \right)\left( {\frac{{L - z}}{D}\theta } \right)\cos (\vartheta )} \right]^{5/6}}dud\vartheta dz \\ 
 \end{array},
 \ee
 where $\theta  = \left| {{{\bm{\theta }}_1} - {{\bm{\theta }}_2}} \right|$ is the angular source separation, $D$ is the aperture diameter, $L$ is the optical path length, and $C_n^2(z)$ is the refractive index structure parameter with $z=0$ at the source.  
 Note that $\left< \cdot \right>$ represents an ensemble mean operator.
 The tilt correlation as a function of separation angle in the direction parallel to the source separation can be shown to be
 \be
 \begin{array}{c}
  \label{parallelcorr}
 {{\tilde r}_\parallel }(\theta ) = \left( { - \frac{{2.91}}{8}} \right){\left( {\frac{{64}}{\pi }} \right)^2}{D^{ - 1/3}}\int\limits_{z = 0}^L {C_n^2\left( z \right)} \int\limits_{\vartheta  = 0}^{2\pi } {\int\limits_{u = 0}^1 {\left[ {\left( {\frac{1}{8}u{{\cos }^{ - 1}}(u)} \right) + u\sqrt {1 - {u^2}} \left( {\frac{1}{{12}}{u^3} - \frac{5}{{24}}u} \right)} \right.} }  \\ 
 \left. { + u\sqrt {1 - {u^2}} \left( {\frac{{{u^3} - u}}{3}} \right){{\cos }^2}(\vartheta )} \right] \times {\left[ {{u^2}{{\left( {\frac{z}{L}} \right)}^2} + {{\left( {\frac{{L - z}}{D}\theta } \right)}^2} + 2u\left( {\frac{z}{L}} \right)\left( {\frac{{L - z}}{D}\theta } \right)\cos (\vartheta )} \right]^{5/6}}dud\vartheta dz \\ 
 \end{array}.
 \ee
Similarly, the tilt correlation as a function of separation angle in the direction perpendicular to the source separation 
is given by
\be
 \label{perpcorr}
\begin{array}{c}
 {{\tilde r}_ \bot }(\theta ) = \left( { - \frac{{2.91}}{8}} \right){\left( {\frac{{64}}{\pi }} \right)^2}{D^{ - 1/3}}\int\limits_{z = 0}^L {C_n^2\left( z \right)} \int\limits_{\vartheta  = 0}^{2\pi } {\int\limits_{u = 0}^1 {\left[ {\left( {\frac{1}{8}u{{\cos }^{ - 1}}(u)} \right) + u\sqrt {1 - {u^2}} \left( {\frac{1}{{12}}{u^3} - \frac{5}{{24}}u} \right)} \right.} }  \\ 
 \left. { + u\sqrt {1 - {u^2}} \left( {\frac{{{u^3} - u}}{3}} \right){{\sin }^2}(\vartheta )} \right] \times {\left[ {{u^2}{{\left( {\frac{z}{L}} \right)}^2} + {{\left( {\frac{{L - z}}{D}\theta } \right)}^2} + 2u\left( {\frac{z}{L}} \right)\left( {\frac{{L - z}}{D}\theta } \right)\cos (\vartheta )} \right]^{5/6}}dud\vartheta dz \\ 
 \end{array}.
\ee
The total tilt correlation in Eq. (\ref{totalcorr}) is related the expressions in Eqs. (\ref{parallelcorr}) and (\ref{perpcorr}) by
\be
{\tilde r_T}(\theta ) = {\tilde r_\parallel }(\theta ) + {\tilde r_ \bot }(\theta ).
\ee

Note that the one-axis Z-tilt variance in units of radians squared is given by  Eqs. (\ref{parallelcorr}) and (\ref{perpcorr}) 
evaluated at a separation angle of 0 yielding
\be
{\tilde \sigma_T}^2={\tilde r_\parallel}(0)={\tilde r_\bot}(0) =  {\tilde r_T }( 0 )/2.
\ee
The tilt correlations can be expressed in terms of separation in pixel spacings 
using the scaling ${r_\parallel }\left( x \right) = {\tilde r_\parallel }\left( {\xi x} \right)/{\xi ^2}$ and
${r_ \bot }\left( x \right) = {\tilde r_ \bot }\left( {\xi x} \right)/{\xi ^2}$, where $\xi$ represents the angle
subtended by one pixel.  Therefore the one-axis Z-tilt variance in units of pixel spacings squared is given by 
\be
{\sigma_T}^2={r_\parallel}(0)={r_\bot}(0) =  {r_T }( 0 )/2.
\ee

\begin{figure}[t]
\begin{center}
{\includegraphics[trim=5cm 9cm 5cm 9cm,scale=.75]{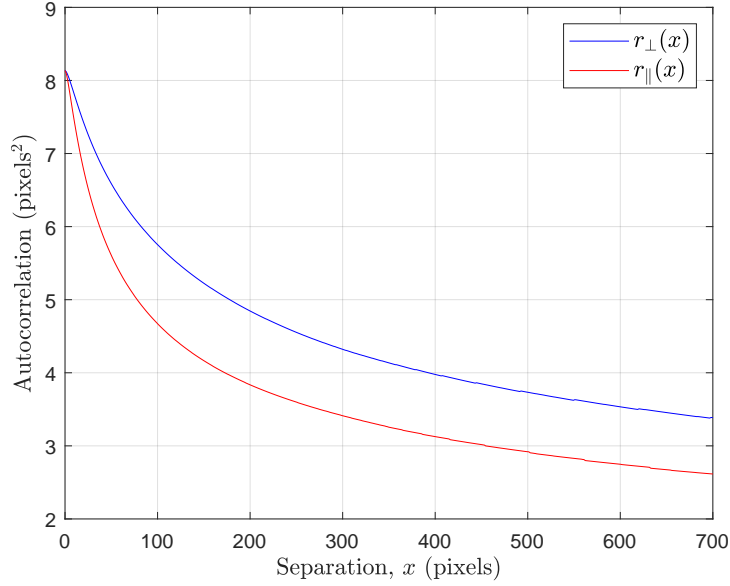}}  
\end{center}
\caption{Tilt correlations for two point sources as a function of perpendicular and parallel separation in pixel spacings
for the optical parameters in Table \ref{optical_parameters} and a constant 
$C_n^2(z) = 1.0\times10^{-15}$ m$^{-2/3}$.}
\label{parper}
\end{figure}

\begin{figure}[t]
\begin{center}
{\includegraphics[trim=0cm 13cm 16cm 0cm,scale=.8]{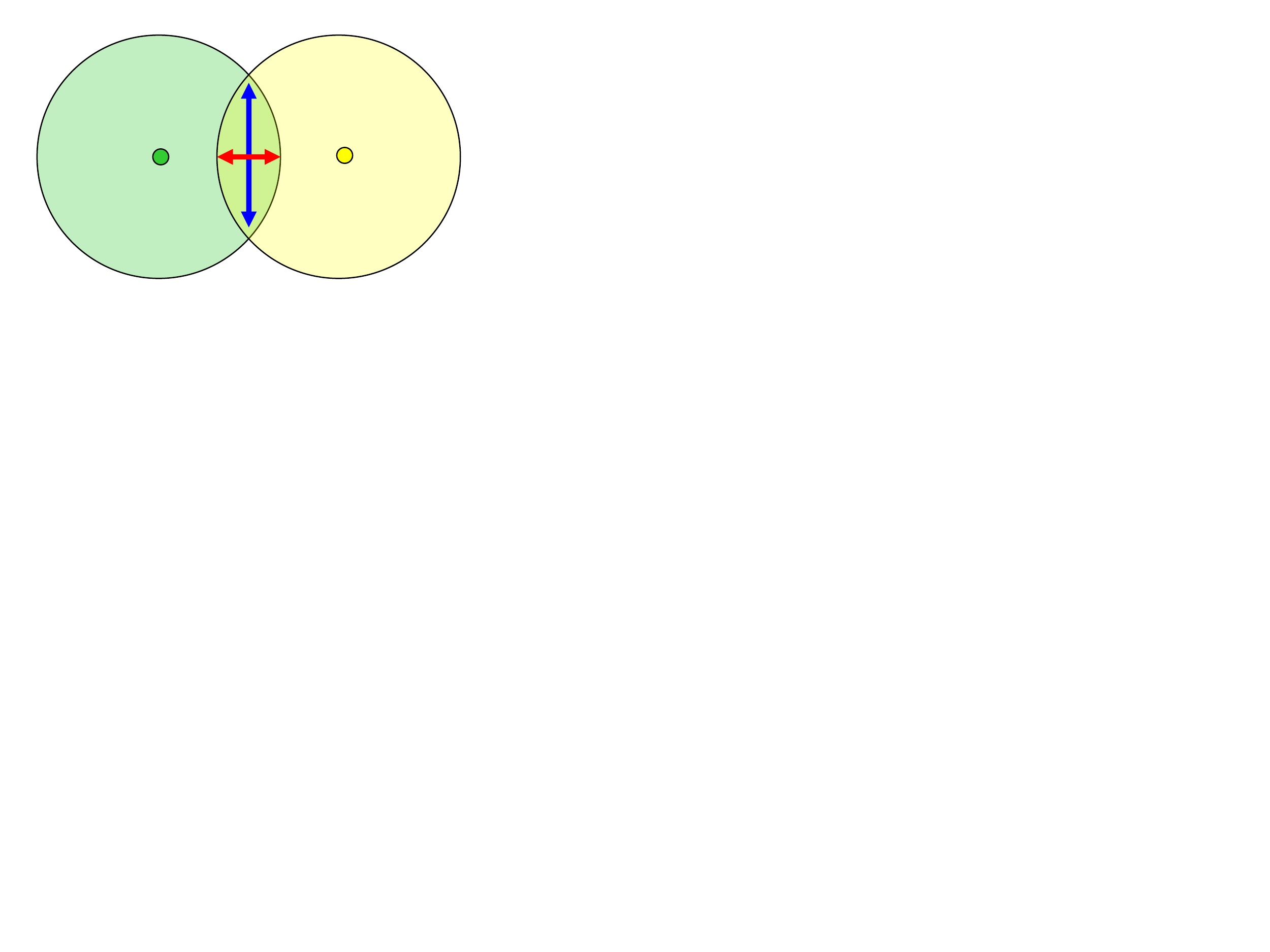}} 
\end{center}
\caption{Isocontours of spherical waves propagating from two point sources at a given distance from the camera. 
The two paths share more of the random medium in common in the perpendicular direction than the parallel direction, giving rise to non-isotropic correlation statistics.}
\label{overlap}
\end{figure}

A pair of the tilt correlation functions is plotted in Fig. \ref{parper} for the optical parameters listed in Table \ref{optical_parameters}
and a constant $C_n^2(z) = 1.0 \times10^{-15}$ m$^{-2/3}$.   
Note that the perpendicular separation shows higher correlation than the parallel  separation.  
To understand this phenomenon, it may be helpful to consider Fig. \ref{overlap}.  Depicted there are
isocontours of spherical waves propagating from two point sources at a given distance from the camera.  Note that 
the two paths share more of the random medium in common in the perpendicular direction than the parallel direction (i.e., the blue arrow is longer than the red).
This is the basis for the asymmetry in the correlation statistics.

\begin{table}[t]
\caption{Optical parameters used for the example and simulation results.} 
\vspace{-.2in}
\label{optical_parameters}
\begin{center}
\begin{tabular}{ | l | l | } 
 \hline
 {\bf Parameter} & {\bf Value} \\
 \hline
Aperture & $D = 0.2034$ m  \\  
Focal length &  $l = 1.2$ m  \\  
F-number &   $f/\# = 5.9$   \\
Wavelength &  $\lambda = 0.525$ $\mu$m  \\
Object distance & $L=7$  km \\ 
Nyquist pixel spacing (focal plane) & $\delta = 1.5488$  $\mu$m \\ 
Nyquist pixel angle & $\xi = 1.2906 \times10^{-6}$ rad \\
 \hline
  \end{tabular}
   \end{center}
\end{table}

\begin{figure}[t]
\begin{center}
{\includegraphics[trim=2cm 11.5cm 18cm 1cm,scale=.9]{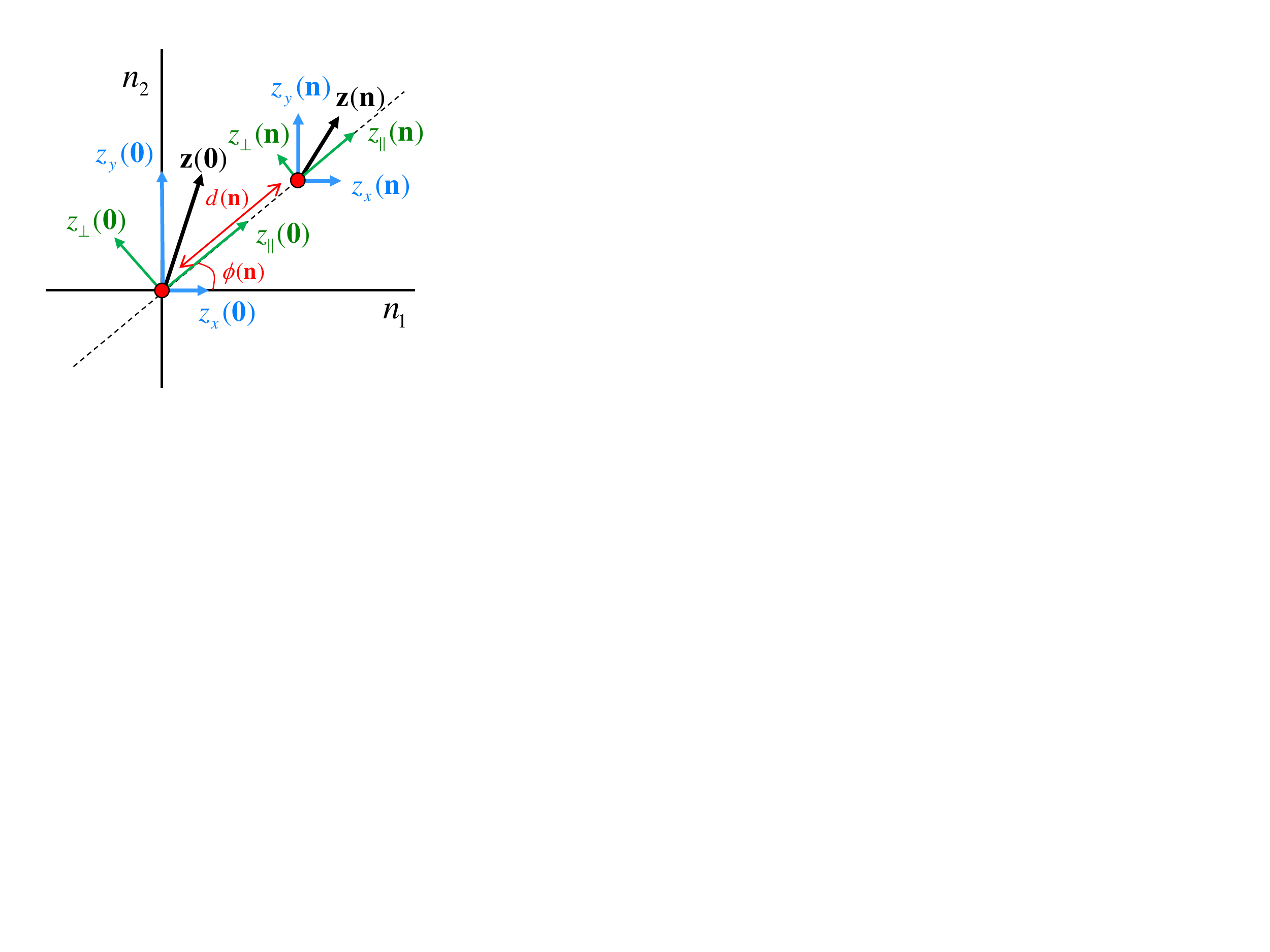}} 
\end{center}
\caption{Geometry for mapping parallel and perpendicular tilt correlations to 2D $x$ and $y$ tilt autocorrelation functions.  Two point sources (red) are shown in the image plane.  The tilt vector are shown as ${\bf z}({\bf 0})$ and ${\bf z}({\bf n})$.  These can be expressed in terms of parallel and perpendicular components (green), or $x$ and $y$ components (blue). }
\label{mapping}
\end{figure}

\subsection{2D Correlation Functions}
\label{corr2dsec}

From a statistical signal processing standpoint, it is more convenient to transform the parallel and perpendicular tilt correlations into more traditional 2D autocorrelation functions, treating them as WSS. 
This can be done in a fashion similar to that presented by Schwartzman {\em et al}~\cite{Schwartzman2017}.  
However, here we use the tilt correlation statistics from Bose-Pillai {\em et al}~\cite{Basu2015,Santasri2018}
that have been validated with our anisoplanatic turbulence simulator~\cite{HardieSimulation2017}.

To derive this mapping, consider the two tilt vectors depicted in Fig. \ref{mapping} in the image plane.    Let the spatial coordinates of the sources be represented in units of pixels.  Without loss of generality, let one of the sources be located at the origin ${\bf 0}=[0,0]^T$ and the other at ${\bf n}=[n_1,n_2]^T$.   These two sources are separated by a distance of 
$d({\bf n})=\sqrt{ n_1^2+n_2^2 }$.  The angle of a line connecting the two sources, relative to the horizontal axis of the 
camera focal plane array, is $\phi({\bf n}) = \tan^{-1}(n_2/n_1)$.
The tilt vectors in the image plane for these two point sources are represented as ${\bf z}({\bf 0})$ and ${\bf z}({\bf n})$, respectively.  These tilt vectors can be expressed in terms of parallel and perpendicular components shown in green, as well as $x$ and $y$ components shown in blue.  
Using a coordinate transformation of the vector components, via vector projection, 
these two forms can be related by 
\be
{z_x}({\bf{n}}) = {z_\parallel }({\bf{n}})\cos \left( {\phi ({\bf{n}})} \right) - {z_ \bot }({\bf{n}})\sin \left( {\phi ({\bf{n}})} \right)
\label{zx}
\ee
and
\be
{z_y}({\bf{n}}) = {z_\parallel }({\bf{n}})\sin \left( {\phi ({\bf{n}})} \right) + {z_ \bot }({\bf{n}})\cos \left( {\phi ({\bf{n}})} \right).
\label{zy}
\ee

\begin{figure}[t]
\begin{center}
{\includegraphics[trim=4cm 6.5cm 5cm 7cm,scale=.75]{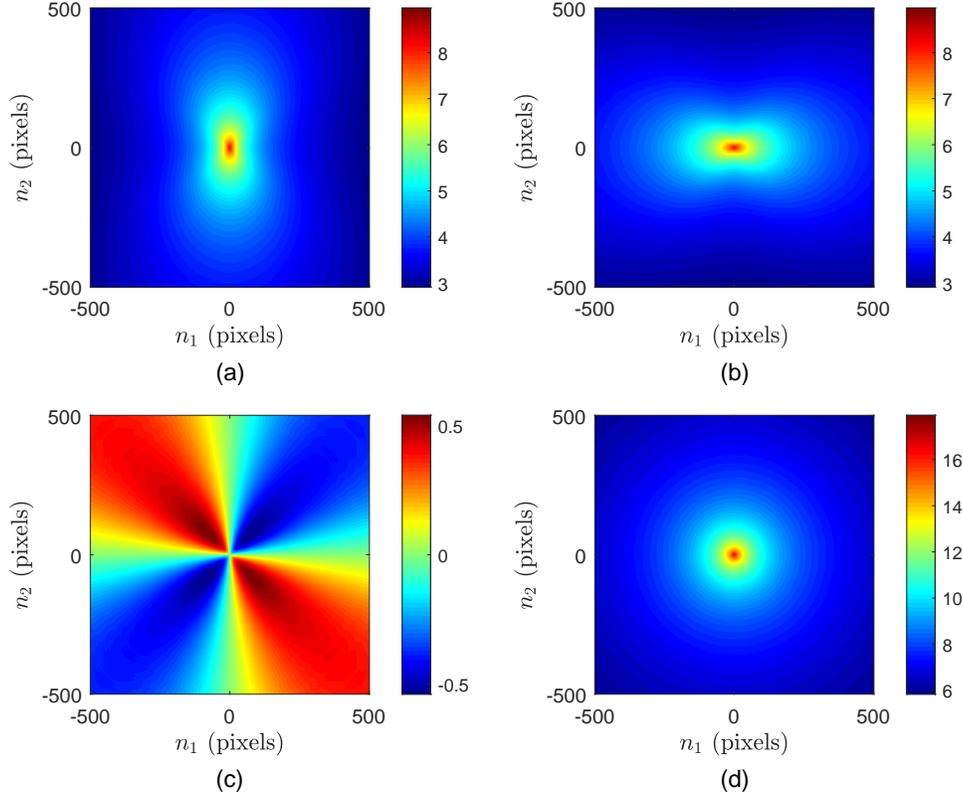}} 
\end{center}
\caption{2D tilt correlation functions corresponding to the statistics shown in Fig. \ref{parper} using 
Eqs. (\ref{rxx}) - (\ref{rxy}).
(a)  ${r_{xx}}({\bf{n}})$, (b)  ${r_{yy}}({\bf{n}})$,   (c)  ${r_{xy}}({\bf{n}}) ={r_{yx}}({\bf{n}})$  and  (d) 
${r_{T}}({\bf{n}}) = {r_{xx}}({\bf{n}}) + {r_{yy}}({\bf{n}})$.  }
\label{corr2dplots}
\end{figure}

Using Eqs. (\ref{zx}) and (\ref{zy}) and noting that the cross-correlation between the parallel and perpendicular components is zero, we obtain the 2D correlation functions shown below.  The 2D autocorrelation for the horizontal axis tilt correlation 
is given by
\be
{r_{xx}}({\bf{n}}) = E\left[ {{z_x}({\bf{0}}){z_x}({\bf{n}})} \right] = {r_\parallel }\left( {d({\bf{n}})} \right){\cos ^2}\left( {\phi ({\bf{n}})} \right) + {r_ \bot }\left( {d({\bf{n}})} \right)\left[ {1 - {{\cos }^2}\left( {\phi ({\bf{n}})} \right)} \right].
\label{rxx}
\ee
Similarly the 2D autocorrelation for the vertical axis tilt correlation 
is given by
\be
{r_{yy}}({\bf{n}}) = E\left[ {{z_y}({\bf{0}}){z_y}({\bf{n}})} \right] = {r_\parallel }\left( {d({\bf{n}})} \right){\sin ^2}\left( {\phi ({\bf{n}})} \right) + {r_ \bot }\left( {d({\bf{n}})} \right)\left[ {1 - {{\sin }^2}\left( {\phi ({\bf{n}})} \right)} \right].
\label{ryy}
\ee
The total of the $x$ and $y$ autocorrelations is the sum
\be
{r_{T}}({\bf{n}})  =  {r_{xx}}({\bf{n}})  + {r_{yy}}({\bf{n}}) = {\tilde r_T }\left( {\xi d({\bf n})  } \right)/{\xi ^2}.
\label{rt}
\ee
Finally, the cross-correlation is given by
\be
{r_{xy}}({\bf{n}}) = {r_{yx}}({\bf{n}}) = E\left[ {{z_x}({\bf{0}}){z_y}({\bf{n}})} \right] = \left( {{r_\parallel }\left( {d({\bf{n}})} \right) - {r_ \bot }\left( {d({\bf{n}})} \right)} \right)\cos \left( {\phi ({\bf{n}})} \right)\sin \left( {\phi ({\bf{n}})} \right).
\label{rxy}
\ee
Applying these transformation to the data in Fig. \ref{parper} produces the 2D correlation functions shown in Fig. \ref{corr2dplots}.   Note that the one-axis Z-tilt variance in units of pixels squared is given by 
$\sigma_T^2 = r_{xx}({\bf 0})=r_{yy}({\bf 0})=  r_{T}({\bf 0})/2$.

\subsection{Tilt Correction Using Image Patch Registration}
\label{residualsec}

Formulating the 2D tilt autocorrelation functions, as done in Section \ref{corr2dsec}, 
allows us to conveniently take advantage of a number of well known statistical signal processing relationships.    
In particular, we are interested in understanding how image registration impacts the 
tilt statistics.  Our idea here is that we can model the impact of image registration 
as filtering random tilt fields with an LSI filter and adding a measurement error term.   Consider 
processing a WSS $x$-dimension random tilt field using an LSI filter with impulse response $h({\bf n})$.
This is denoted as
\be
{{\bar z}_x}({\bf n}) = {z_x}({\bf n})*h({\bf n}) + {e_x}({\bf n}),
\ee
where ${e_x}({\bf n})$ is the measurement error associated with the $x$ tilt at position ${\bf n}$.
Let us treat the measurement error as zero-mean independent and identically distributed white noise with
standard deviation $\sigma_e$ in units of pixel spacings.   In this case, the output autocorrelation can be expressed as
\be
{\bar r_{xx}}({\bf{n}}) = {r_{xx}}({\bf{n}})*h({\bf{n}})*h( - {\bf{n}}) + \sigma_e^2\delta({\bf{n}}),
\label{rxxconv}
\ee
or equivalently
\be
{\bar r_{xx}}({\bf{n}}) = \sigma_e^2\delta({\bf{n}}) + \sum\limits_{\bf{m}}^{} {\sum\limits_{\bf{k}}^{} {h({\bf{m}})h({\bf{k}})} } {r_{xx}}({\bf{n}} + {\bf{m}} - {\bf{k}},{\bf{n}} + {\bf{m}} - {\bf{k}}),
\label{rxxsum}
\ee
where ${\bf m}=[m_1,m_2]^T$, ${\bf k}=[k_1,k_2]^T$, and $\delta({\bf n})$ is a 2D  Kronecker delta function.
The variance of the processed random tilt field is given by
\be
\bar \sigma _{xx}^2 = {\bar r_{xx}}({\bf{0}}) = \sigma_e^2+ \sum\limits_{\bf{m}}^{} {\sum\limits_{\bf{k}}^{} {h({\bf{m}})h({\bf{k}})} } {r_{xx}}({\bf{m}} - {\bf{k}},{\bf{m}} - {\bf{k}}).
\label{varsum}
\ee
Of course, the same can be done for the $y-$dimension tilt to produce
${\bar r_{yy}}({\bf n})$ and $\bar \sigma _{yy}^2$ from ${r_{yy}}({\bf n})$.
Note that if the filter impulse response  is isotropic in $x$ and $y$, the tilt variance 
result will be the same whether you use 
 $r_{xx}({\bf n})$, $r_{yy}({\bf n})$, or ${r_{T}}({\bf{n}}) /2$.  That is, in the isotropic case,
$\bar \sigma _{yy}^2$ = $\bar \sigma _{xx}^2$.  
For non-isotropic filters, the impact of the filtering on the $x$ and $y$ tilt variance must be computed separately.

Now let us consider the special
case of averaging the tilts over a square patch with side dimension of $2M+1$.  This is equivalent to
applying a moving average filter to the tilt fields with impulse response
\be
h_{P}({\bf n}) = \left\{ {\begin{array}{*{20}{c}}
   {\frac{1}{{{{\left( {2M + 1} \right)}^2}}}} & {|{n_1}|,|{n_2}| \le M{\rm{ }}}  \\
   0 & {{\rm{otherwise}}}  \\
\end{array}} \right..
\label{mafilter}
\ee
Using this impulse response plugged into Eqs. (\ref{rxxconv}) or (\ref{rxxsum}) and including measurement error 
gives the output tilt correlation function for estimated patch tilt.  It can be applied to $r_{xx}({\bf n})$, $r_{yy}({\bf n})$, or ${r_{T}}({\bf{n}})$.
Combining Eqs. (\ref{varsum}) and (\ref{mafilter}) gives the variance of estimated square-patch tilt as
\be
\sigma _P^2 = \sigma_e^2+\frac{1}{{{{(2M + 1)}^4}}}\sum\limits_{{\bf{m}} =  - M{\bf{1}}}^{M{\bf{1}}} {\sum\limits_{{\bf{k}} =  - M{\bf{1}}}^{M{\bf{1}}} {{r_{xx}}({\bf{m}} - {\bf{k}},{\bf{m}} - {\bf{k}})} },
\label{mavar}
\ee
where ${\bf 1}=[1,1]^T$.   Because the moving average filter is isotropic, one could equivalently use
$r_{xx}({\bf n})$, $r_{yy}({\bf n})$, or ${r_{T}}({\bf{n}}) /2$ in Eq. (\ref{mavar}).

\begin{figure}[t]
\begin{center}
{\includegraphics[trim=4cm 6.5cm 5cm 7cm,scale=.8]{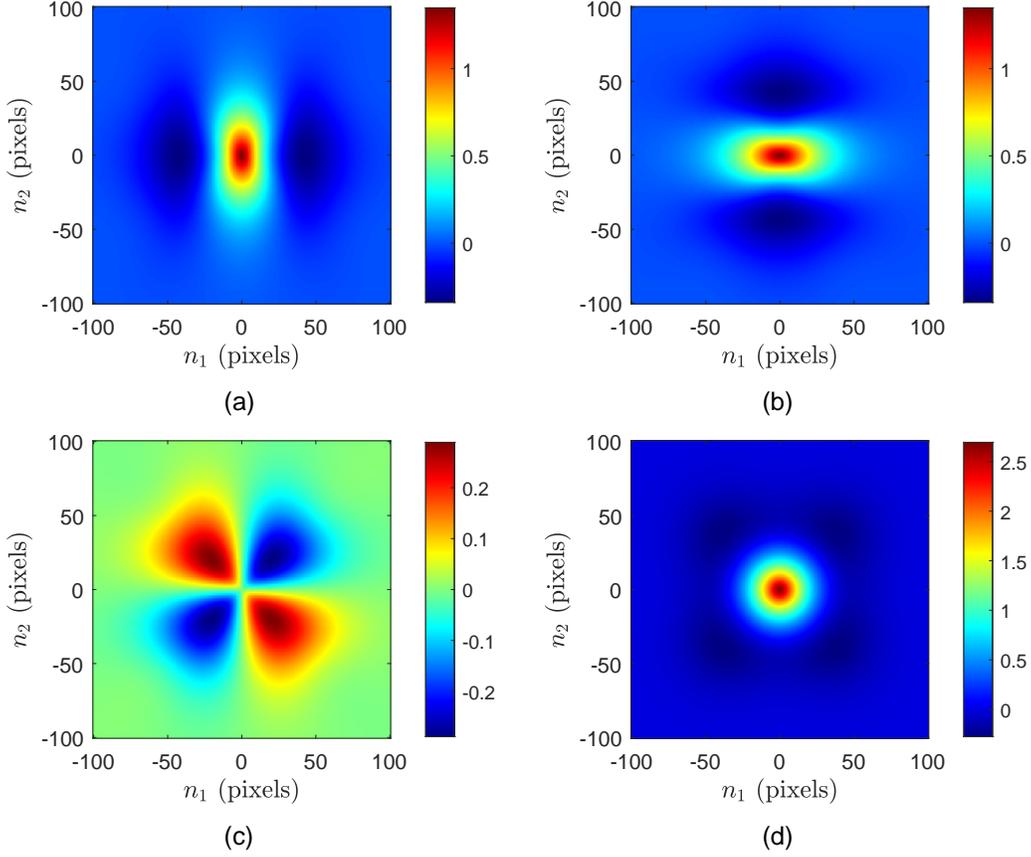}} 
\end{center}
\caption{2D residual tilt correlation functions corresponding to those in Fig. \ref{corr2dplots} after filtering the tilt fields using the impulse response in Eq. (\ref{bmafilter}) for $M=50$ (i.e., $101\times101$ patch tilt correction).
(a)  ${{\bar r}_{xx}}({\bf{n}})$, (b)  ${{\bar r}_{yy}}({\bf{n}})$,   (c)  ${{\bar r}_{xy}}({\bf{n}}) ={{\bar r}_{yx}}({\bf{n}})$  and  (d) 
${{\bar r}_{T}}({\bf{n}}) = {{\bar r}_{xx}}({\bf{n}}) + {{\bar r}_{yy}}({\bf{n}})$.  }
\label{filteredcorr2dplots}
\end{figure}

To model the patch-based image registration that attempts to correct for 
these patch tilts, we propose using an identity filter minus a moving average filter.
The output of this difference filter represents the residual tilt after registration.
The residual tilt impulse response for a $2M+1$ square patch size is
\be
h_{R}({\bf n})=\delta({\bf n})-h_{P}({\bf n}).
\label{bmafilter}
\ee
Using this impulse response plugged into Eqs. (\ref{rxxconv}) or (\ref{rxxsum}) gives the output tilt correlation function
after the patch registration.  As with the patch tilt correlations, 
we can obtain the post-registration residual tilt statistics corresponding to
 $r_{xx}({\bf n})$, $r_{yy}({\bf n})$, or ${r_{T}}({\bf{n}})$.
Examples of the correlation functions from Fig. \ref{corr2dplots} after filtering 
with $h_{R}({\bf n})$ for $M=50$ (i.e., $101\times101$ patch tilt correction) are shown in Fig. \ref{filteredcorr2dplots} for
$\sigma_e=0$.
Looking at the vertical colorbar scales, one can see that the residual tilt variance is greatly reduced
compared with  Fig. \ref{corr2dplots} as a result of such a registration process.

\begin{figure}[t]
\begin{center}
{\includegraphics[trim=5cm 9cm 5cm 9cm,scale=.75]{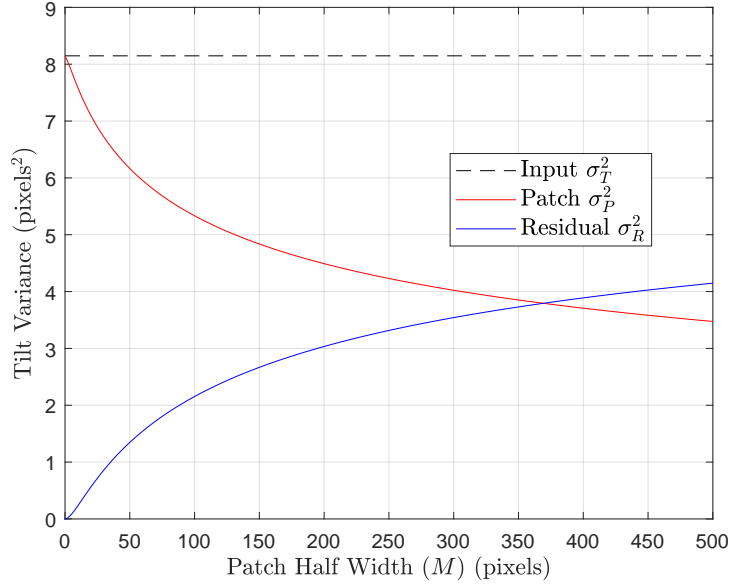}} 
\end{center}
\caption{Tilt variances as a function of patch half width $M$ for the optical parameters in Table \ref{optical_parameters} and $C_n^2(z) = 1.0\times10^{-15}$ m$^{-2/3}$.  Shown are the constant input tilt variance $\sigma_T^2$, the patch tilt variance
$\sigma_P^2$,  and the residual tilt variance $\sigma_R^2$.}
\label{patchtiltvar}
\end{figure}

The residual tilt variance can then be computed as a function of block size using the impulse response
in Eq. (\ref{bmafilter}) and the relationship in Eq. (\ref{varsum}) yielding 
\be
\sigma _R^2 = \sigma_e^2+\sum\limits_{{\bf{m}} =  - M{\bf{1}}}^{M{\bf{1}}} {\sum\limits_{{\bf{k}} =  - M{\bf{1}}}^{M{\bf{1}}} {{h_R}({\bf{m}}){h_R}({\bf{k}}){r_{xx}}({\bf{m}} - {\bf{k}},{\bf{m}} - {\bf{k}})} } .
\label{sigmaR2}
\ee
The patch tilt variance, $\sigma_P^2$, and residual tilt variance, $\sigma_R^2$,
are plotted in Fig. \ref{patchtiltvar} as a function of the square patch half width $M$ for
$\sigma_e=0$.
These correspond to the optical parameters in Table \ref{optical_parameters} and $C_n^2(z) = 1.0\times10^{-15}$ m$^{-2/3}$.
 Also shown is the original unprocessed tilt variance $\sigma_T^2$.  Note that as the patch size increases, the
patch tilt variance decreases, and the residual tilt variance increases. 
We can quantify the level of tilt variance reduction accomplished by the patch registration by comparing the 
residual tilt variance $\sigma _{R}^2$ to the total input tilt variance $\sigma _T^2$.
In particular, we define the tilt correction factor as
\be
\alpha  = 1 - \frac{{\sigma _R^2}}{{\sigma _T^2}} = 1 - \varepsilon  - \frac{1}{{{r_{xx}}({\bf{0}})}}\sum\limits_{{\bf{m}} =  - M{\bf{1}}}^{M{\bf{1}}} {\sum\limits_{{\bf{k}} =  - M{\bf{1}}}^{M{\bf{1}}} {{h_R}({\bf{m}}){h_R}({\bf{k}}){r_{xx}}({\bf{m}} - {\bf{k}},{\bf{m}} - {\bf{k}})} },
\label{alphaMlong}
\ee
where $\varepsilon  = \sigma _e^2/\sigma _T^2$ is the registration error-to-signal ratio.
Again, note that we use the relationship $\sigma_T^2 = r_{xx}({\bf 0})$ in expressing Eq. (\ref{alphaMlong}). 
Also, note that the autocorrelation function $ r_{xx}({\bf n})$ 
is derived from the correlations in  Eqs. (\ref{totalcorr}) - (\ref{perpcorr}).   
These correlations are a function of $C_n^2(z)$.  However, when $C_n^2(z)$ is constant, it can be brought out of all of 
the integrals.    Since the rightmost term in Eq. (\ref{alphaMlong}) has 
$r_{xx}(\cdot)$ in the numerator and denominator, this constant will cancel.
If we further assume $\varepsilon$ is constant, then
$\alpha$ is not a function of turbulence strength.
This means that for constant $C_n^2(z)$, the tilt correction factor is only a function of camera parameters, 
optical path length, and the patch size $M$.  Thus, in this case we can compute $\alpha$ without {\em apriori} knowledge of the
turbulence strength.

Note that the tilt correction factor $\alpha$ generally ranges from 0 to 1.  However, it is possible for this parameter to become negative
if the residual tilt variance, $\sigma_R^2$, is larger than the input tilt variance, $\sigma_T^2$.  
This would imply that error in the registration is larger than any
tilt correction being performed, resulting in a net increase in tilt variance.   
It would be advisable in this scenario to forgo such ineffective registration if the goal is to reduce tilt variance.

\begin{figure}[t]
\begin{center}
{\includegraphics[trim=5cm 9cm 5cm 9cm,scale=.75]{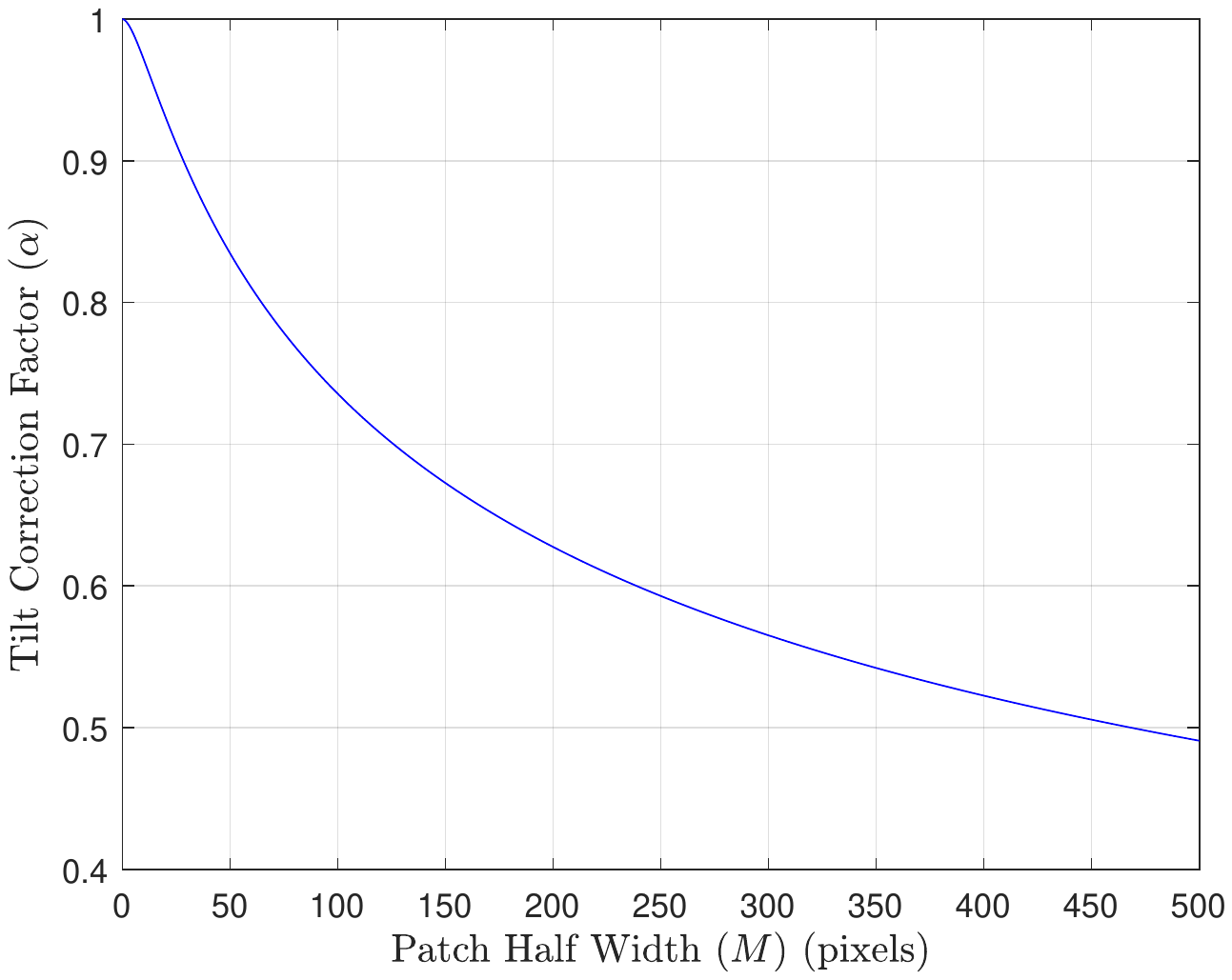}} 
\end{center}
\caption{Tilt correction factor $\alpha$ as a function of $M$ for the optical parameters in Table \ref{optical_parameters}, $\varepsilon=0$, and any constant $C_n^2$.   Using a larger registration patch provides less tilt correction, but
larger patches can yield a more accurate motion estimate.}
\label{alphaMfig}
\end{figure}

The tilt correction factor for the parameters in Table \ref{optical_parameters} and 
$C_n^2(z) = 1.0\times10^{-15}$ m$^{-2/3}$  is plotted as a function of $M$ in
Fig. \ref{alphaMfig} for $\varepsilon =0 $.  
For large $M$, only large scale low-frequency tilts are corrected, 
giving rise to a relatively small tilt correction factor.  
However, large patch tilts can be estimated much more accurately than smaller ones because of the number of
image pixels involved.   As $M$ decreases, the turbulence warping compensation is occurring on a smaller spatial scale, leaving 
less residual tilt variance, and generating more tilt correction.

\begin{figure}[tbh]
\begin{center}
{\includegraphics[trim=5cm 9cm 5cm 9cm,scale=.75]{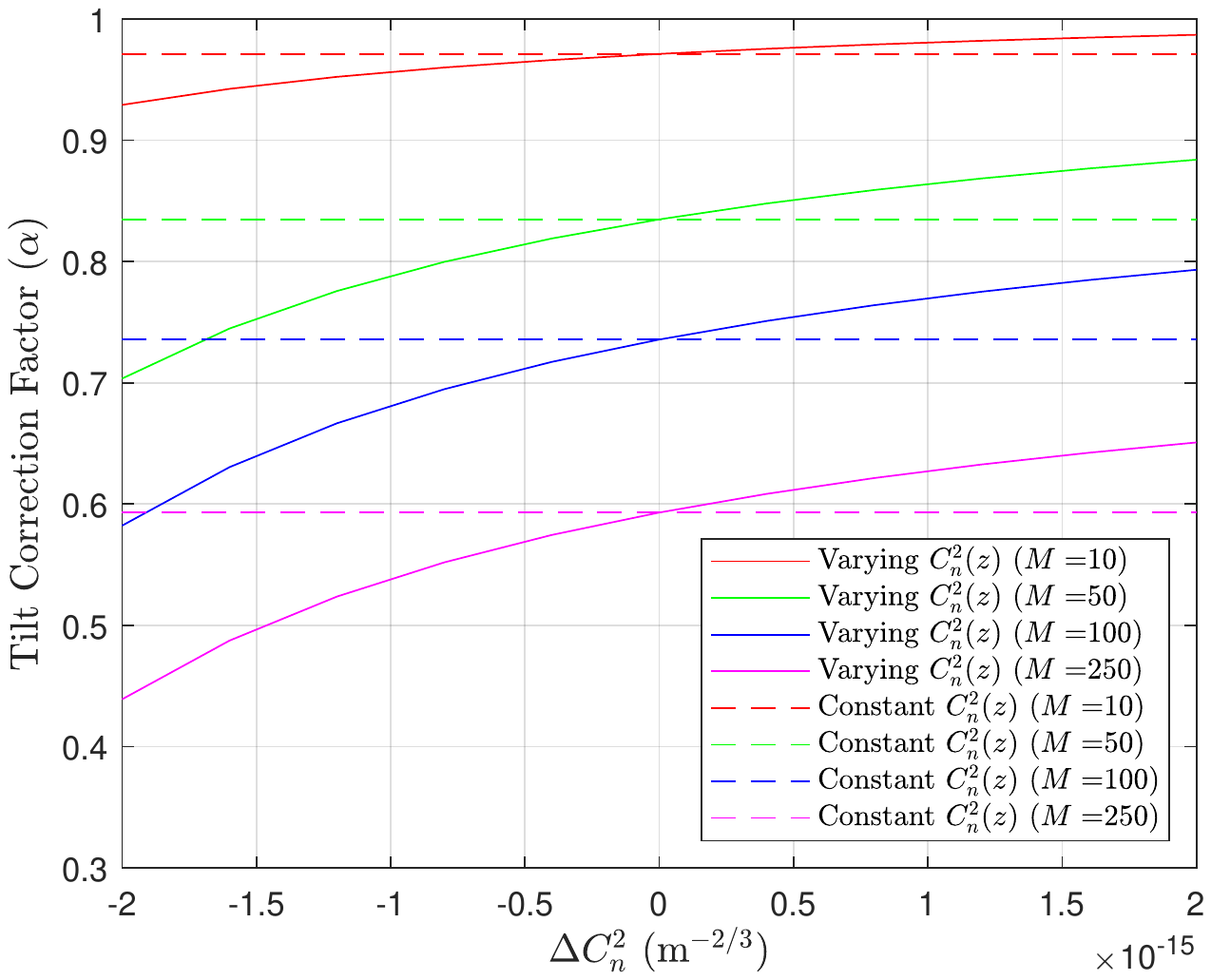}} 
\end{center}
\caption{Tilt correction factor $\alpha$ as a function of the source-to-camera change for a linear $C_n^2(z)$ profile using 
the optical parameters in Table \ref{optical_parameters} and $\varepsilon=0$.   The path average value is 
$C_n^2=1\times10^{-15}$ m$^{-2/3}$. }
\label{alpha_sensitivity_fig}
\end{figure}

We have also conducted a sensitivity analysis to see the impact of a non-constant $C_n^2(z)$ profile on $\alpha$.   
The results are summarize in Fig. \ref{alpha_sensitivity_fig}.
In this analysis we consider a linear $C_n^2(z)$ profile with an average value of 
$C_n^2=1\times10^{-15}$ m$^{-2/3}$ and 
source-to-camera change of $\Delta C_n^2$.   
Shown in Fig. \ref{alpha_sensitivity_fig} are the true $\alpha$ values for 
the varying $C_n^2(z)$ path for four values of $M$ with $\varepsilon =0$.
Also shown are the corresponding values obtained assuming a constant $C_n^2$.
Note that for $\Delta C_n^2=-2\times10^{-15}$, we have strong turbulence at the
source, and no turbulence at the camera.  In this case, the true path-dependent $\alpha$ is lower than what is predicted 
using a constant path assumption.  This means the registration will be less effective in tilt correction than the constant path 
prediction.  Object-heavy turbulence gives rise to a lower isoplanatic angle and more localized warping~\cite{HardieSimulation2017}.  Thus, it stands to reason that block-based registration would be less effective in this scenario.
Note that the sensitivity illustrated in Fig. \ref{alpha_sensitivity_fig} is reduced with smaller values of $M$.

\subsection{Tilt Correction with Global Registration}
\label{globalsec}

In the previous section, we considered patch-based image registration such as 
the BMA for tilt reduction.  
However, it is also interesting to apply this tilt autocorrelation function analysis to the scenario of global image registration.  
The global registration effectively seeks to determine the average tilt across the full image and then correct all pixels with this one global shift.
Thus, imagine a patch filter $h_{P}({\bf n})$ where $2M+1$ spans the full image size.  Such a filter outputs 
the global average tilt shifts.
The correction filter would no longer be spatially invariant as we used in Eq. (\ref{bmafilter}).  Rather, the residual filter would
vary with each pixel's location ${\bf k}=[k_1,k_2]^T$, yielding
\be
h_R({\bf n}; {\bf k})=\delta({\bf n}+{\bf k})-h_{P}({\bf n}) .
\label{globalfilter}
\ee

The residual tilt variance can now be computed in a fashion similar to that in Eq. (\ref{sigmaR2}).  Because the filter in Eq. (\ref{globalfilter}) is
non-isotropic, the residual tilt variance in $x$ and $y$ must be computed separately using the corresponding tilt correlation function.
Furthermore,  the residual tilt variance must be computed separately for each pixel ${\bf k}$.
We can use the residual tilt variance along with the total tilt variance to form a tilt correction factor similar to that in
Eq. (\ref{alphaMlong}).  However, now the tilt correction factor is a function of pixel position and is different in $x$ and $y$.

\begin{figure}[t]
\begin{center}
{\includegraphics[trim=5cm 10cm 5cm 10cm,scale=.75]{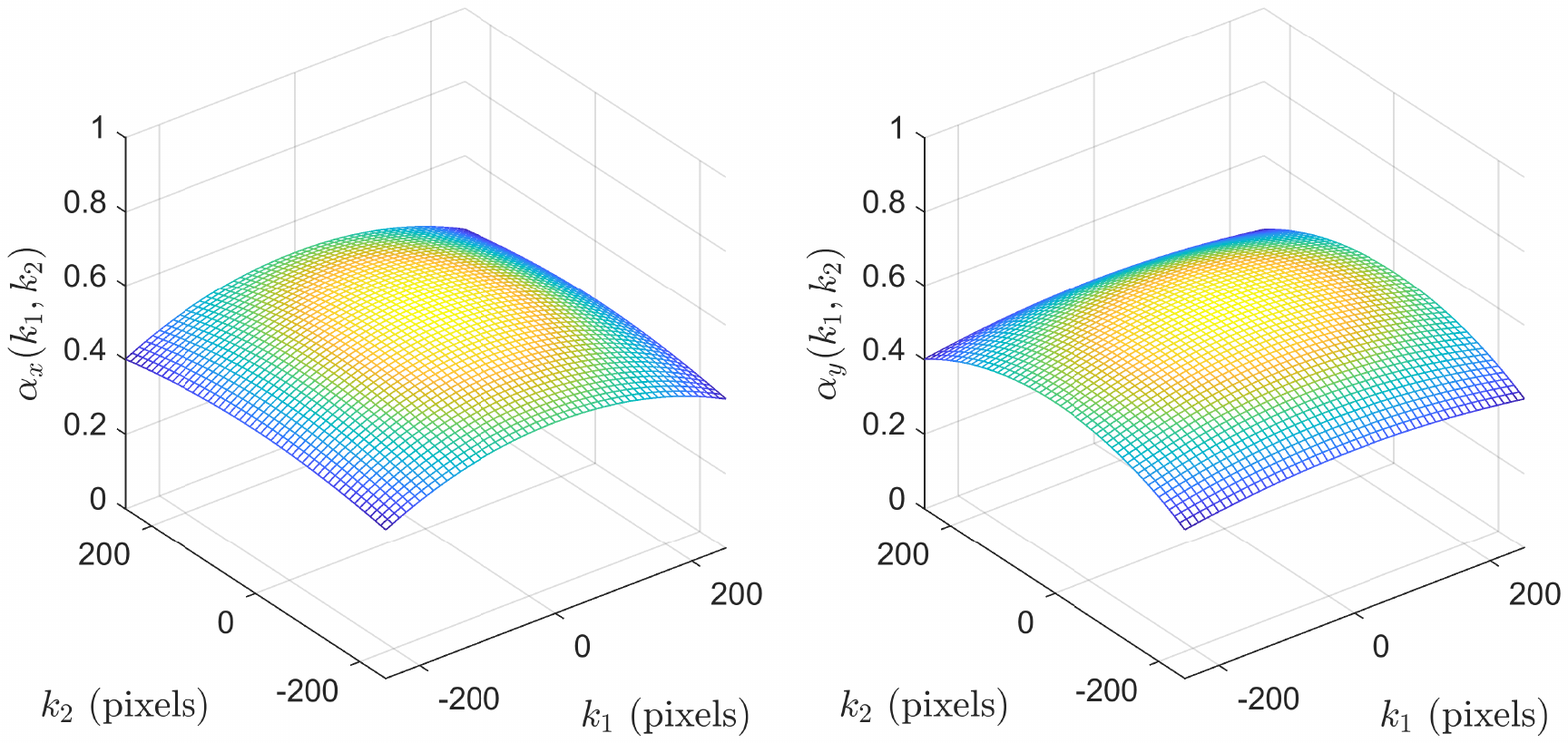}} 
\end{center}
\caption{Global registration tilt correction factors as a function of ${\bf k} = [k_1,k_2]^T$ with $M=250$ (i.e., $501\times501$ image), $\varepsilon=0$, and using the optical parameters in Table \ref{optical_parameters}.  
The $x$-dimension function is on the left and the 
$y$-dimension function is on the right.  }
\label{global_reg_corr}
\end{figure}

These global tilt correction factor functions are plotted in Fig. \ref{global_reg_corr} for
$M=250$ with the optical parameters in Table \ref{optical_parameters}.  Note that there is more tilt correction in the center of the image and less in the corners.  This behavior is explained by the fact that global registration will be most representative of the bulk of the image that is contained in the center.   The global shifts are least representative of the corners.  
Notwithstanding this spatially varying relationship, we have observed that
useful results in our parameter estimation problems can be achieved by using the average tilt correction factor and neglecting the
spatial variation.  Note in Fig. \ref{global_reg_corr} that the peak value of the two surfaces is 0.5903 which matches Fig. \ref{alphaMfig} for $M=250$.  The tilt reduction factor obtained by averaging over all pixels is $\alpha = 0.5181$.

\section{Application to Turbulence OTF Modeling}
\label{otfappsec}

Let us now turn our attention to defining the OTF in the observation model introduced in Eqs. (\ref{halphaz}) and (\ref{Halphaz}).
In our model we shall include the effects of diffraction-limited optics, the average short exposure atmospheric OTF, and the averaging of frames with residual tilt variance\cite{bmwf2017,FIFSR2019}.
The OTF combining these components is given by
\be
H_\alpha(\rho ) = H_{\rm{dif}}(\rho ) H_{\rm{SE}}(\rho)G_{\alpha}(\rho),
\label{Halpha2}
\ee
where $H_{\rm dif}(\rho)$ is for diffraction-limited optics, $H_{\rm SE}(\rho)$ is the average short-exposure OTF, and
$G_{\alpha}(\rho)$ captures blurring from averaging frames with residual tilt variance.
These component functions are circularly symmetric and the radial frequency parameter $\rho  = \sqrt {{u^2} + {v^2}}$, 
where $u$ and $v$ are the spatial frequencies in units of cycles per unit distance in the camera focal plane. 

The diffraction-limited OTF for a circular exit pupil\cite{GOODOPT} is given by
\be
{H_{\rm{dif}}}(\rho ) = \left\{ {\begin{array}{*{20}{c}}
   {\frac{2}{\pi }\left[ {{{\cos }^{ - 1}}\left( {\frac{\rho }{{2{\rho _c}}}} \right) - \frac{\rho }{{2{\rho _c}}}\sqrt {1 - {{\left( {\frac{\rho }{{2{\rho _c}}}} \right)}^2}} } \right]} & {\rho  \le {\rho _c}}  \\
   0 & {{\rm{otherwise}}}  \\
\end{array}} \right.,
\label{Hdif}
\ee
where ${\rho _c} = 1/(\lambda f/\# )$ is the optical cut-off frequency, and the f-number is $f/\#  = l/D$.
For the average short exposure OTF, we will use Fried's near-field model based on Kolmogorov statistics~\cite{FRIED1966,ROGWEL1996}. 
This is given by 
\be
H_{\rm{SE}}(\rho ) = \exp \left\{ { - 3.44{{\left( {\frac{{\lambda l\rho }}{{{r_0}}}} \right)}^{5/3}}\left[ {1 - {{\left( {\frac{{\lambda l\rho }}{D}} \right)}^{1/3}}} \right]} \right\}.
\label{HSE}
\ee
Fried~\cite{FRIED1966} states that the near-field approximation is for $D \gg \sqrt{L\lambda}$.  However, Tofsted\cite{Tofsted2003,Tofsted2010,Tofsted2011} showed that
this approximation is quite good even up to $D \approx \sqrt{L\lambda}$. 

The term ${G_\alpha}(\rho ) $ is a Gaussian function that corresponds to blurring from residual tilt variance.
We wish to define this function so that when
$\alpha=0$ (no tilt correction) we have
\be
H_{\rm{SE}}(\rho )G_{0}(\rho)= H_{\rm{LE}}(\rho ) ,
\ee
where $H_{\rm{LE}}(\rho )$ is Fried's long exposure OTF model~\cite{FRIED1966,ROGWEL1996} given by
\be
H_{\rm{LE}}(\rho ) = \exp \left\{ - 3.44{{\left( {\frac{{\lambda l\rho }}{{{r_0}}}} \right)}^{5/3}} \right\}.
\label{HLE}
\ee
Also, for $\alpha=1$ (perfect tilt correction), we want $G_{1}(\rho)=1$ so that
\be
H_{\rm{SE}}(\rho )G_{1}(\rho)= H_{\rm{SE}}(\rho ).
\ee
As shown by Hardie {\em et al}\cite{bmwf2017}, we can achieve this using the relationship 
\be
{G_\alpha }(\rho ) = \exp \left\{   - \frac{3.44 (1-\alpha)(\lambda l \rho)^2} {r_0^{5/3}D^{1/3} }    \right\}.
\ee
 This has the Gaussian form
 \be
{G_\alpha }(\rho ) = \exp \left\{ {\frac{{ - {\rho ^2}}}{{2\sigma _G^2(\alpha)}}} \right\},
\label{Galpha}
\ee
where
   \be
\sigma_G^2(\alpha) = \frac{{r_0}^{5/3}{D^{1/3}}}{6.88{ ( 1 - \alpha ) {{\left( {\lambda l} \right)}^2}}}.
\label{sigmaG2r0}
\ee

In the spatial domain, Eq. (\ref{Halpha2}) may be expressed as
\be
h_\alpha(r) =  h_{\rm{dif}}(r)*h_{\rm{SE}}(r)*g_{\alpha}(r),
\ee
where $r=\sqrt{x^2+y^2}$ is the radial spatial variable.
The term $h_{\rm{dif}}(r)$ is the inverse Fourier transform of Eq. (\ref{Hdif}), $h_{\rm{SE}}(r)$ 
is the inverse Fourier transform of Eq. (\ref{HSE}), and ${g_\alpha }(r)$ is a spatial domain Gaussian\cite{bmwf2017}.
This Gaussian is given by
\be
{g_\alpha }(r) = \frac{{2\sigma _g^2(\alpha)}}{\pi }\exp \left\{ { - \frac{{{r^2}}}{{2\sigma _g^2(\alpha)}}} \right\},
\label{galpha}
\ee
where
\be
\sigma _g^2(\alpha ) =\frac{1}{4\pi^2\sigma_G^2(\alpha)}.
\label{sigma_g2}
\ee

\begin{figure}[t]
\begin{center}
{\includegraphics[trim=5cm 9cm 6cm 9cm,scale=.67]{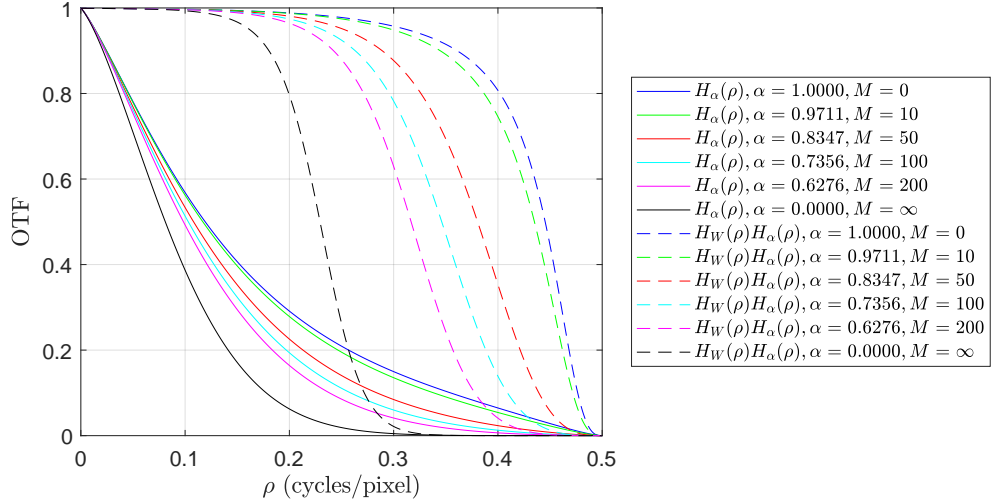}} 
\end{center}
\caption{Overall OTF plots from Eq. (\ref{Halpha2}) for the Optical parameters in Table \ref{optical_parameters} and 
$C_n^2(z) = 2.5\times10^{-16}$ m$^{-2/3}$.   Also shown are the OTFs with the Wiener filter
from Eq. (\ref{wiener}) applied.   The different curves are for different patch sizes $M$ assuming $\varepsilon=0$.  As $M$ increases, the tilt correction factor decreases, residual blurring increases, and the OTF becomes more low-pass in nature.  }
\label{wienerotf}
\end{figure}

Some example OTFs are shown in Fig. \ref{wienerotf} using the optical parameters in Table \ref{optical_parameters} and 
$C_n^2(z) = 2.5\times10^{-16}$ m$^{-2/3}$.   The different curves are for different values of $M$ with $\varepsilon=0$.  
As $M$ increases, the tilt correction factor decreases for anisoplanatic turbulence.   In turn, the residual blurring increases, 
and the OTF becomes more low-pass in nature.
Note that in the limit when $M=0$ ($1\times1$ block), 
we have ideal tilt correction with $\alpha=1$, and the resulting OTF is the average atmospheric 
short-exposure OTF.  On the other extreme, as $M$ approaches infinity, $\alpha$ approaches 0 giving us the long exposure OTF.
Also shown in Fig. \ref{wienerotf} as dashed lines are the Wiener filter compensated OTFs for $\Gamma=0.001$.  
These curves include the Wiener OTF from Eq. (\ref{wiener}) as a function of the radial frequency variable $\rho$.
Here the differences in the OTFs are even more pronounced.  
The OTF curves in Fig. \ref{wienerotf} suggest that the smallest $M$ should be used to produce the most tilt correction and
the most favorable OTF.  However, in practice, as the 
patch size is decreased the accuracy of the motion estimates declines and requires a higher $\varepsilon$.  Thus, registration
for OTF improvement requires balancing the patch size $M$ with the motion estimation error $\varepsilon$ to create the highest $\alpha$ possible.

\section{Spectral Ratio Fried Parameter Estimation}
\label{ratio_section}

We shall now introduce a modified spectral ratio method for estimating the Fried parameter, $r_0$, that can be applied to registered images to address camera motion.   
This method involves acquiring a sequence of short exposure frames as one would do for image restoration for turbulence mitigation.  We will use the short exposure images and also 
form a long exposure image by averaging the frames.
The proposed method involves dividing the spatial-frequency spectrum of 
long-exposure image by that of the short-exposure images~\cite{vonderLuhe:84}.
Based on our model in Section \ref{otfappsec}, the result should have a Gaussian form.
The Fried parameter can be found as a function of the Gaussian variance.
One of the novel aspects of the approach presented here lies in how utilize the tilt
correction factor to compensate for image registration.

Consider a pristine image spectrum $Z(u,v)$.  Acquiring this image with diffraction-limited optics and average short-exposure OTF, we obtain the spectrum
\be
{Y_{1}}(u,v) = {H_{1}}(\rho )Z(u,v) =  H_{\rm{dif}}(\rho ) H_{\rm{SE}}(\rho ) G_1(\rho ) Z(u,v).
\label{Y1}
\ee
Note that we are using $\alpha=1$ because this is assumed to be a true average short-exposure image.
Now consider forming a modified long exposure image by averaging the short exposure frames.
Let this be given by
\be
{Y_{\alpha}}(u,v) = {H_{\alpha}}(\rho )Z(u,v) = H_{\rm{dif}}(\rho ) H_{\rm{SE}}(\rho )  G_\alpha(\rho ) Z(u,v).
\label{Yalpha}
\ee
The reason we do not set $\alpha=0$ is that some registration to compensate for camera motion
might be necessary.   Compensating for camera motion will also inadvertently compensate for
some turbulence motion.   Using the $\alpha$ from Eq. (\ref{alphaMlong}) gives us 
the means to account for this warping tilt reduction as a function of the registration window size.
If we form a spectral ratio from Eqs. (\ref{Yalpha}) and (\ref{Y1}), 
we obtain
\be
\frac{{{Y_{\alpha }}(u,v)}}{{{Y_{1}}(u,v)}} = 
\frac{H_\alpha(\rho ) Z(u,v) }{H_1(\rho) Z(u,v)} =
\frac{H_{\rm{dif}}(\rho ) H_{\rm{SE}}(\rho )  G_\alpha(\rho ) Z(u,v)}
{ H_{\rm{dif}}(\rho ) H_{\rm{SE}}(\rho ) G_1(\rho )  Z(u,v)}.
\ee
Canceling the like terms and recognizing that $G_1(\rho ) = 1$ yields
\be
\frac{{{Y_{\alpha }}(u,v)}}{{{Y_{1}}(u,v)}} = 
\frac{H_\alpha(\rho ) }{H_1(\rho)} = G_\alpha(\rho ).
\ee
Thus, the spectral ratio has a Gaussian form given by Eq. (\ref{Galpha}) with variance $\sigma_G^2(\alpha)$ 
given by Eq. (\ref{sigmaG2r0}).  Let an image-based estimate of this variance be denoted $\hat{\sigma}_G^2$.
Finally, using our estimated Gaussian spectral ratio variance in Eq. (\ref{sigmaG2r0}) and solving for $r_0$ 
we obtain the Fried parameter estimate
\be
{\hat r_0} = {\left[ {\frac{{{\rm{ }} 6.88 {{\left( {\lambda l{{\hat \sigma }_G}} \right)}^2}({1 - \alpha } )}}{{{D^{1/3}}}}} \right]^{3/5}}.
\label{r0sigmaG}
\ee
Thus, the problem of $r_0$ estimation is transformed to that of estimating $\sigma_G^2(\alpha)$ 
from the spectral ratio function.  
The parameter $\alpha$ in Eq. (\ref{r0sigmaG}) gives us the ability to compensate for any
camera-motion registration performed.

\begin{figure}[t]
\begin{center}
{\includegraphics[trim=5cm 9cm 5cm 9cm,scale=.7]{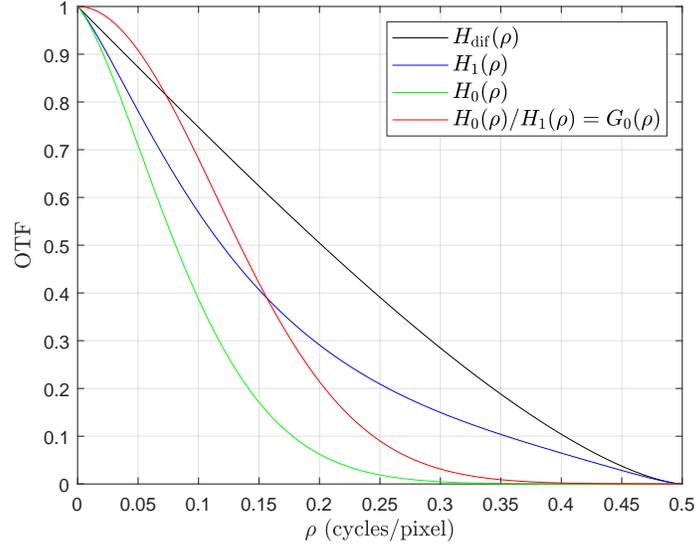}} 
\end{center}
\caption{OTF plots showing the diffraction-limited OTF, short-exposure OTF, long-exposure OTF, and the
spectral ratio of the long exposure over the short exposure OTF for the optical parameters in Table \ref{optical_parameters} and $C_n^2(z) = 2.5\times10^{-16}$ m$^{-2/3}$.  The ratio is a Gaussian and the standard deviation is related to the 
Fried parameter by Eq. (\ref{r0sigmaG}).}
\label{specratioplot}
\end{figure}

Example spectra are shown in Fig. \ref{specratioplot} to illustrate the Gaussian nature of the spectral ratio.   
 The optical parameters are those in Table \ref{optical_parameters}
and $C_n^2(z) = 2.5\times10^{-16}$ m$^{-2/3}$.  Shown are the diffraction limited OTF, $H_{\rm dif}(\rho)$,
the short-exposure OTF with diffraction, $H_1(\rho)$, the long exposure OTF with diffraction, $H_0(\rho)$, 
and the ratio that is equal to $G_0(\rho)$.   Note that the ratio does have a Gaussian appearance 
as expressed in Eq. (\ref{Galpha}).   When registration is necessary to compensate for camera motion, the long exposure 
OTF is replaced with $H_\alpha(\rho)$ and the ratio is given by $G_\alpha(\rho)$ for $\alpha \in (0,1)$.   
Note that as $\alpha$ gets larger and approaches 1, the short- and modified long-exposure OTFs
converge.  This makes estimating the Fried parameter very sensitive to noise and other errors.  
To prevent this, we recommend using the largest $M$ possible for $r_0$ estimation.  
Alternatively, one may use global registration and apply the average $\alpha$ as described in Section \ref{globalsec}.
For restoration purposes, smaller $M$ is desirable, so long as the tilt motion can be estimated accurately.
Thus, we propose a two-step process where a large $M$ is first used to compensate for camera motion and estimate $r_0$.  In a subsequent step, the images would be re-registered using BMA with
a relatively small $M$ to provide maximum tilt correction and give the most favorable OTF for the best restoration.

\begin{figure}[t]
\begin{center}
{\includegraphics[trim=1cm 10cm 5cm 1cm,scale=.67]{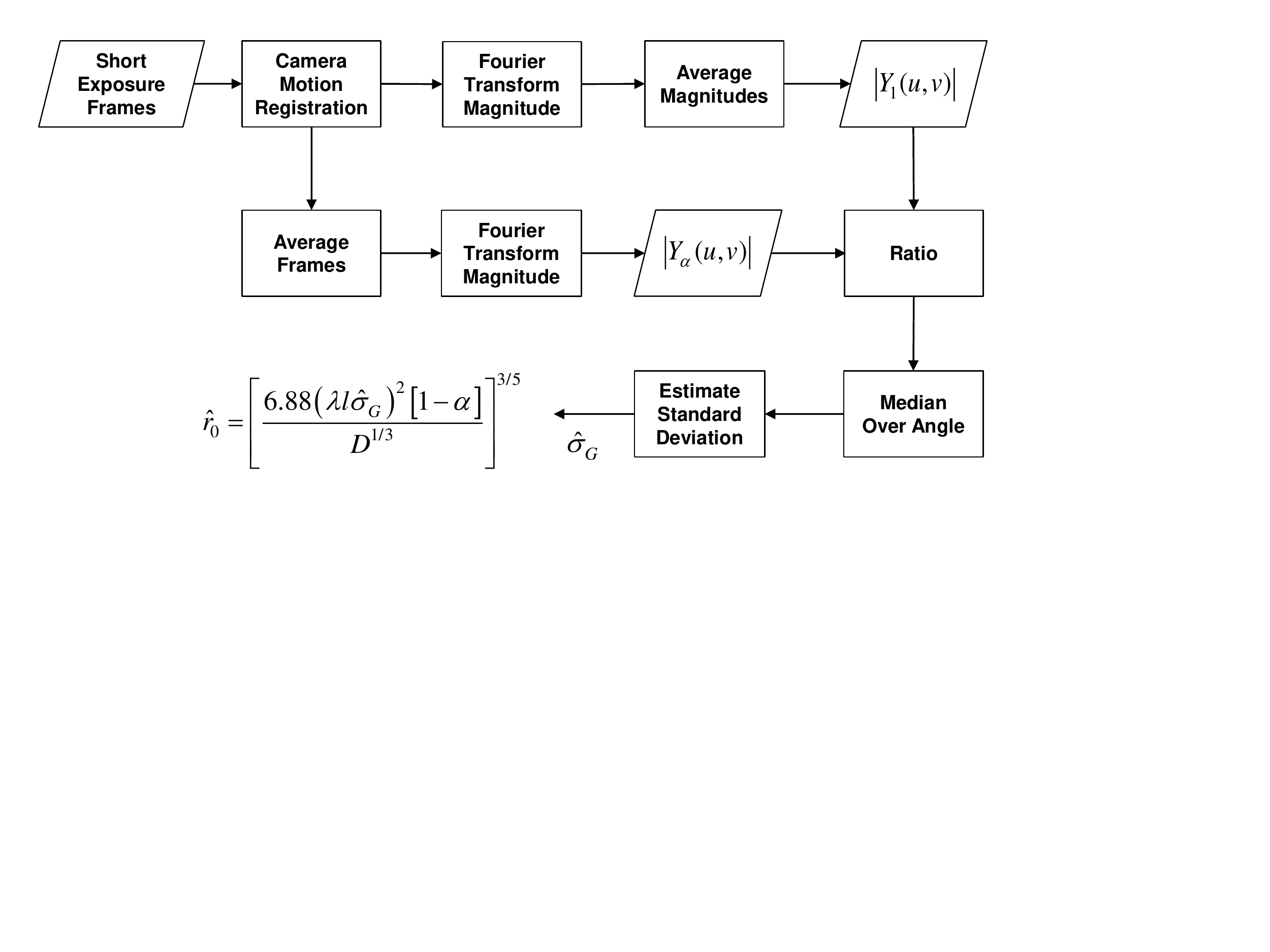}} 
\end{center}
\caption{Block diagram of the proposed spectral ratio Fried parameter estimation method with camera motion registration and compensation using the tilt correction factor $\alpha$.}
\label{specratioblock}
\end{figure}

A block diagram of the proposed spectral ratio Fried parameter estimation method is presented in Fig. \ref{specratioblock}.
The process begins with the acquisition of short-exposure frames.  If camera motion is present, global registration 
or BMA registration using the largest practical block size is performed.   
The registration is important when camera motion is present because we do not want camera motion to produce additional non-turbulence motion blurring that would  impact the Fried parameter estimate.  
At the same time, we want only the minimum turbulence tilt compensation when forming the long exposure image.  
Unfortunately, by correcting for camera motion, we will invariably also be correcting for some large-scale 
turbulence warping as well.
Turbulence tilt correction is kept to a minimum when we use the largest $M$ possible, as shown in Fig. \ref{alphaMfig}.
Global registration offers the best option for correcting camera motion with minimal turbulence tilt correction.

Based on the type of registration, if any, we must generate a corresponding $\alpha$ that represents the level of turbulence
tilt correction.  If there is no camera motion, no registration is done and we simply set $\alpha=0$.   
If BMA registration is used, we set $\alpha$ using Eq. (\ref{alphaMlong}) based on the block size governed by $M$.
If global registration is used, we use the $\alpha$ computed according to Section \ref{globalsec} based on the full image size.

The registered (or raw) frames are then averaged to form a camera-motion-compensated long-exposure image. 
Next, the magnitude spatial-frequency spectrum of the long-exposure image is computed.
Also, the magnitude spectra of the individual short-exposure frames are computed and averaged.  
The 2D spectral ratio array is then formed and converted into polar coordinates.  
A robust estimate of the underlying radial function 
is formed by computing the median across all angles for each radial distance.  
Curve fitting is applied to estimate the Gaussian standard deviation $\hat{\sigma}_G$.  Finally,
Eq. (\ref{r0sigmaG}) is used to form the estimate $\hat{r}_0$.  Note that since we have discrete space images, 
the Fourier transforms are computed using the FFT.   
Also, windowing with a Tukey window is applied before all FFTs to reduce border discontinuity effects.

A desirable characteristic of the proposed method is that all pixels in the image are utilized 
in forming the $r_0$ estimate, adding to robustness.   
In addition, if image registration is used to address camera motion,
the impact of this registration may be accounted for with the tilt correction factor $\alpha$.  
As mentioned earlier, this parameter may be determined in the case of
constant $C_n^2(z)$ based on optical parameters and the block or image size described by $M$.   
Results for specific registration methods can be fine-tuned using the registration error-to-signal ratio parameter $\varepsilon$ in Eq. (\ref{alphaMlong}).  We have obtained excellent results using $\varepsilon = 1/12$ with integer pixel BMA.  
Note that this is the variance of quantization noise for integer quantization.   

One last point to note is that the proposed algorithm assumes a static scene so that the long- and short-exposure images
correspond to one another in terms of underlying scene content.  In the case of moving objects within the scene, the ``movers''
will be present in the short-exposure imagery but will be blurred away in the long exposure imagery.  
This will tend to have the impact of reducing the $\hat{\sigma}_G$ and producing an 
erroneously low $\hat{r}_0$ estimate.  Thus, it would be helpful to detect scene motion regions~\cite{VanHook2020} and avoid 
these for $r_0$ estimation.

\section{Experimental Results}
\label{results_section}

In this section we present a number of experimental results to demonstrate the efficacy of
the proposed spectral ratio $r_0$ estimation algorithm presented in Section \ref{ratio_section} 
and turbulence mitigation using the OTF model in Section \ref{otfappsec}.  We first present results using
simulated data in Section \ref{simsec}
that allow for a detailed quantitative performance analysis.  Next, we use real data in Section \ref{realsec}.

\subsection{Simulated Data}
\label{simsec}

\begin{figure}[t]
\begin{center}
{\includegraphics[trim=2cm 10cm 2cm 9cm,scale=.75]{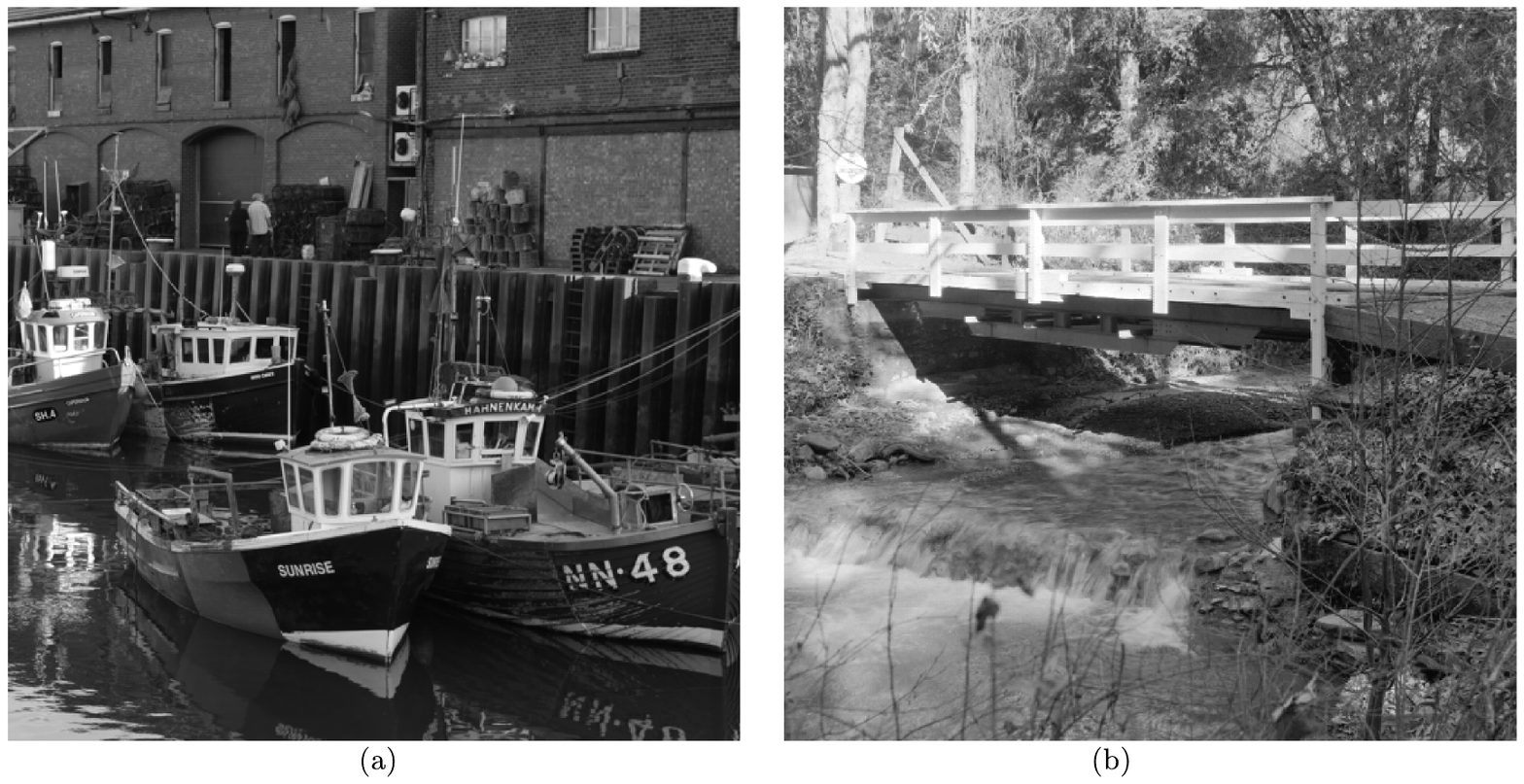}} 
\end{center}
\caption{Standard publicly available truth images used for simulation results.  The images are converted to grayscale and resized to a pixel size of  $501\times501$. (a) Boats~\cite{Agustsson_2017_CVPR_Workshops},  (b) Stream and Bridge~\cite{usc_images}.}
\label{truth_images_fig}
\end{figure}

The simulated data are generated using the numerical wave-propagation tool developed by Hardie 
{\em et al}~\cite{HardieSimulation2017}.
This simulator produces realistic anisoplanatic turbulence degradations and it 
has been validated using a number of
key turbulence statistics~\cite{HardieSimulation2017}.  We use the optical parameters in 
Table \ref{optical_parameters} and the turbulence parameters
listed in Table \ref{turbulence_parameters}.   The detailed simulator parameters match those
in the original simulator paper~\cite{HardieSimulation2017}.
All of the simulated frames are derived from the truth images shown in Fig. \ref{truth_images_fig}.  
These are standard publicly available 8-bit grayscale images sized to $501\times501$ pixels.  
As can be seen in Table \ref{turbulence_parameters}, we model 6 turbulence levels with constant $C_n^2(z)$ profiles.
For each truth image and each turbulence level, 300 temporally independent frames are generated.  Additive Gaussian noise is
included with a standard deviation of 1 digital unit.
Table \ref{turbulence_parameters} shows several statistics for each turbulence level.

\begin{table}[t] 
\small
\caption{Turbulence parameters for 6 levels of turbulence generated with the anisoplanatic turbulence simulator\cite{HardieSimulation2017}
to test the spectral ratio $r_0$ estimator.  The patch and residual tilt variance values are for $M=100$ ($201\times201$ pixel patch).}
\vspace{-.2in}
\label{turbulence_parameters}
\begin{center}
 \begin{tabular}{ |l|c|c|c|c|c|c| }  
\hline
& \multicolumn{6}{c|}{{\bf Turbulence Degradation}} \\
\cline{2-7}
{\bf Parameter} & {\bf Level 1} & {\bf Level 2} & {\bf Level 3}  & {\bf Level 4} & {\bf Level 5} & {\bf Level 6}  \\ 
 \hline \hline 
$C_n^2 \times 10^{-15}$ (m$^{-2/3})$ & 0.1000 & 0.2500 & 0.5000  & 1.0000  & 1.5000 & 2.0000  \\  \hline  
Theoretical $r_0$ (m) & 0.1901 & 0.1097 & 0.0724  &  0.0478  & 0.0374 & 0.0315  \\ \hline  
Theoretical $D/r_0$ (unitless)& 1.0697 & 1.8536 & 2.8096  &  4.2585  & 5.4314 & 6.4547  \\ \hline  
Isoplanatic Angle (pixels)& 6.6174 & 3.8188 & 2.5194  &  1.6622  & 1.3033 & 1.0966  \\ \hline  
RMS Tilt (pixels)& 0.9026 & 1.4272 & 2.0183  &  2.8543  & 3.4958 & 4.0367  \\ \hline  
Tilt Variance (pixels$^2$)& 0.8147 & 2.0368 & 4.0736  &  8.1473  & 12.2209 & 16.2946  \\ \hline  
Patch Tilt Variance (pixels$^2$)& 0.5333 & 1.3333 & 2.6666  &  5.3333  & 7.9999 & 10.6666  \\ \hline  
Residual Tilt Variance (pixels$^2$)& 0.2154 & 0.5385 & 1.0770  &  2.1541  & 3.2311 & 4.3082  \\ \hline  
 \end{tabular}
 \end{center}
\end{table}

\subsubsection{Fried Parameter Estimation with Simulated Data}

\begin{figure}[t]
\begin{center}
{\includegraphics[trim=5cm 9cm 5cm 9cm,scale=.75]{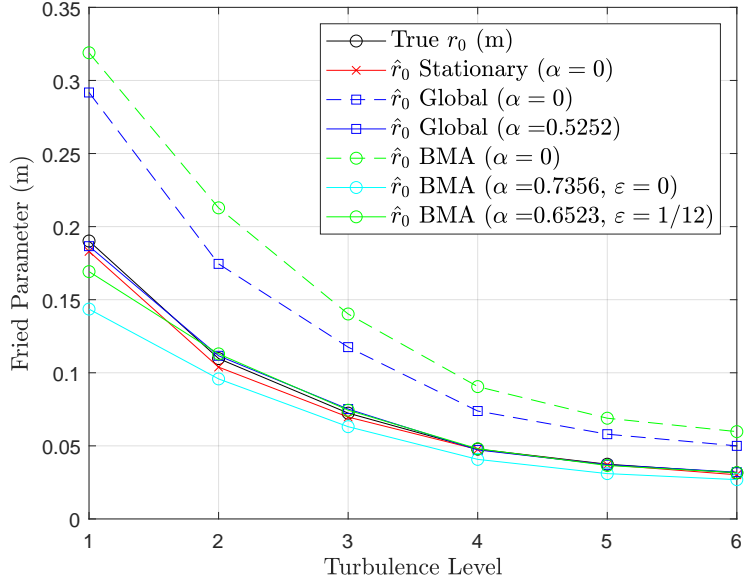}} 
\end{center}
\caption{Spectral ratio Fried parameter estimates for the 6 levels of turbulence in Table \ref{turbulence_parameters}
and the image Boats.}
\label{r0est_plot}
\end{figure}

For each turbulence level, we use several variations of our spectral ratio method to estimate the Fried parameter.  The
results are reported in Table \ref{r0_error_table_boats} for the Boats image and Table \ref{r0_error_table_stream} for Stream and Bridge.  The results for Boats are also plotted in Fig. \ref{r0est_plot} for the reader's convenience.  The percent errors for the various estimates compared with the true $r_0$ are also shown in the tables. 

The estimates labeled ``Stationary'' use no registration and would be appropriate for a stationary camera.  Since there is no camera motion, we use $\alpha=0$, as there is no tilt correction taking place by means of registration.  This method of estimating $r_0$ 
has a maximum absolute percent error of 5.34\% for the datasets used.  Now consider the case of using global registration to compensate for possible camera motion.  If this is done without applying an appropriate tilt correction factor, the $r_0$ estimate is significantly inflated.  This is because the camera registration compensates for turbulence motion, making the turbulence seem weaker, giving rise to a high Fried parameter estimate. This effect can be seen clearly 
in Fig. \ref{r0est_plot} for for the curve labeled ``Global ($\alpha=0$)''.  
However, by computing and applying the global tilt correction
factor, as described in Section \ref{globalsec}, excellent results are achieved.  This can be seen 
in Fig. \ref{r0est_plot} for for the curve labeled ``Global ($\alpha=0.5252$)''.    
This method has a maximum absolute percent error of 5.71\% for the datasets used.  

The remaining estimates are formed using BMA registration with $M=100$ ($201\times201$ patch size).
As expected, the uncompensated BMA estimate (i.e., $\alpha=0$) is too large.  Interestingly, using the $\alpha$
as computed in Eq. (\ref{alphaMlong}) with $\varepsilon=0$, we get an underestimate of the Fried parameter.  
This is due to the fact that BMA 
registration is imperfect and is computed here only to the nearest whole pixel.   Thus, it is achieving less
tilt correction than a theoretically ideal patch registration of that size.  By setting the registration error-to-signal value to be
$\varepsilon=1/12$, this effect can be largely compensated for, as shown in Fig. \ref{r0est_plot}.

It should be noted that the global registration is a more practical method for compensating for
camera motion.  It can be done with subpixel accuracy very efficiently.  We employ a normalized cross-correlation
to get to the nearest whole pixel, and then follow this with the iterative gradient method of Lucas and Kanade~\cite{Lucas1981} for subpixel accuracy.  Because this global registration is very accurate, we use $\varepsilon=0$. The larger effective block size of global registration also allows for less turbulent tilt correction (i.e., smaller $\alpha$), 
and this generally helps to improve the $r_0$ estimate.  
In contrast, BMA with large block sizes is very computationally demanding, especially if one seeks subpixel accuracy.  
The maximum block size with BMA is also more limited because of border effects.

\begin{table}[p] 
\small
\caption{Spectral ratio Fried parameter estimation error analysis for the 6 levels of turbulence detailed in Table \ref{turbulence_parameters} and the truth image Boats.}
\vspace{-.2in}
\label{r0_error_table_boats}
\begin{center}
 \begin{tabular}{ |l|c|c|c|c|c|c| }  
\hline
& \multicolumn{6}{c|}{{\bf Turbulence Degradation}} \\
\cline{2-7}
{\bf Parameter} & {\bf Level 1} & {\bf Level 2} & {\bf Level 3}  & {\bf Level 4} & {\bf Level 5} & {\bf Level 6}  \\ 
 \hline \hline 
Theoretical $r_0$ (m) & 0.1901 & 0.1097 & 0.0724  &  0.0478  & 0.0374 & 0.0315  \\ \hline  
Stationary $\hat{r}_0$ (m) & 0.1830 & 0.1039 & 0.0697  &  0.0475  & 0.0371 & 0.0302  \\  
Percent Error  & -3.75\% & -5.34\% & -3.73\%  &  -0.46\%  & -1.01\% & -4.17\%  \\ \hline  
Global $\hat{r}_0$ (m) ($\alpha=0$) & 0.2913 & 0.1738 & 0.1168  &  0.0734  & 0.0575 & 0.0495  \\   
Percent Error  & 53.20\% & 58.37\% & 61.40\%  &  53.66\%  & 53.66\% & 57.23\%  \\ \hline  
Global $\hat{r}_0$ (m) ($\alpha=0.5252$) & 0.1863 & 0.1112 & 0.0747  &  0.0469  & 0.0368 & 0.0317  \\   
Percent Error  & -2.01\% & 1.30\% & 3.24\%  &  -1.72\%  & -1.71\% & 0.57\%  \\ \hline  
BMA $\hat{r}_0$ (m) ($\alpha=0$) & 0.3189 & 0.2128 & 0.1402  &  0.0905  & 0.0689 & 0.0597  \\   
Percent Error  & 67.74\% & 93.96\% & 93.69\%  &  89.55\%  & 84.04\% & 89.56\%  \\ \hline  
BMA $\hat{r}_0$ (m) ($\varepsilon=0$) & 0.1436 & 0.0958 & 0.0631  &  0.0408  & 0.0310 & 0.0269  \\  
Percent Error & -24.50\% & -12.69\% & -12.81\%  &  -14.68\% & -17.16\% & -14.67\%  \\ \hline  
BMA $\hat{r}_0$ (m) ($\varepsilon=1/12$) & 0.1692 & 0.1129 & 0.0744  &  0.0480  & 0.0366 & 0.0317  \\  
Percent Error & -11.01\% & 2.91\% & 2.77\%  &  0.57\% & -2.35\% & 0.57\%  \\ \hline  
 \end{tabular}
\end{center}

\vspace{.2in}

\small
\caption{Spectral ratio Fried parameter estimation error analysis for the 6 levels of turbulence detailed in Table \ref{turbulence_parameters} and the truth image Stream and Bridge.}
\vspace{-.1in}
\label{r0_error_table_stream}
\begin{center}
 \begin{tabular}{ |l|c|c|c|c|c|c| }  
\hline
& \multicolumn{6}{c|}{{\bf Turbulence Degradation}} \\
\cline{2-7}
{\bf Parameter} & {\bf Level 1} & {\bf Level 2} & {\bf Level 3}  & {\bf Level 4} & {\bf Level 5} & {\bf Level 6}  \\ 
 \hline \hline 
 Theoretical $r_0$ (m) & 0.1901 & 0.1097 & 0.0724  &  0.0478  & 0.0374 & 0.0315  \\ \hline  
Stationary $\hat{r}_0$ (m) & 0.1823 & 0.1058 & 0.0692  &  0.0473  & 0.0371 & 0.0298  \\  
Percent Error  & -4.15\% & -3.56\% & -4.41\%  &  -0.99\%  & -1.00\% & -5.33\%  \\ \hline  
Global $\hat{r}_0$ (m) ($\alpha=0$) & 0.2803 & 0.1684 & 0.1121  &  0.0716  & 0.0556 & 0.0466  \\   
Percent Error  & 47.41\% & 53.50\% & 54.80\%  &  49.94\%  & 48.42\% & 47.80\%  \\ \hline  
Global $\hat{r}_0$ (m) ($\alpha=0.5252$) & 0.1793 & 0.1077 & 0.0717  &  0.0458  & 0.0356 & 0.0298  \\   
Percent Error  & -5.71\% & -1.81\% & -0.98\%  &  -4.10\%  & -5.07\% & -5.46\%  \\ \hline  
BMA $\hat{r}_0$ (m) ($\alpha=0$) & 0.3096 & 0.2071 & 0.1415  &  0.0899  & 0.0663 & 0.0565  \\   
Percent Error  & 62.81\% & 88.71\% & 95.41\%  &  88.24\%  & 77.13\% & 79.42\%  \\ \hline  
BMA $\hat{r}_0$ (m) ($\varepsilon=0$) & 0.1394 & 0.0932 & 0.0637  &  0.0405  & 0.0299 & 0.0254  \\  
Percent Error & -26.71\% & -15.05\% & -12.04\%  &  -15.27\% & -20.27\% & -19.24\%  \\ \hline  
BMA $\hat{r}_0$ (m) ($\varepsilon=1/12$) & 0.1643 & 0.1099 & 0.0751  &  0.0477  & 0.0352 & 0.0300  \\  
Percent Error & -13.62\% & 0.12\% & 3.68\%  &  -0.13\% & -6.02\% & -4.81\%  \\ \hline 
 \end{tabular}
 \end{center}
\end{table}

\begin{figure}[t]
\begin{center}
{\includegraphics[trim=5cm 9cm 5cm 8cm,width=5.25in]{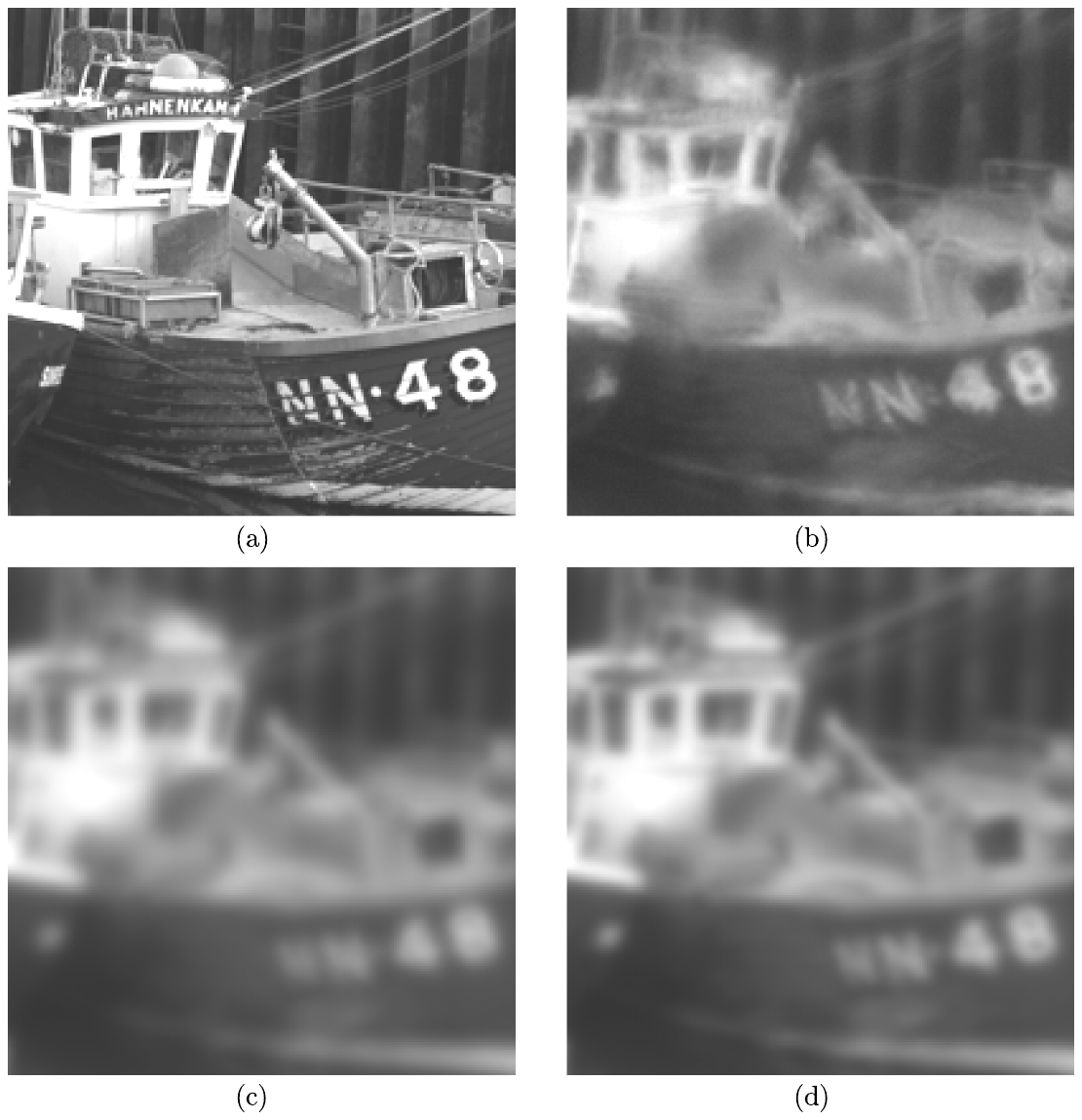}} 
\end{center}
\caption{Region of interest images for level 4 turbulence from Table \ref{turbulence_parameters} for the image Boats.  (a)  Truth image, (b) Frame 1 short-exposure image, (c) long-exposure image with no camera motion, (d) motion compensated frame average using global registration.}
\label{boat_roi}
\end{figure}

To better understand the impact of registration on the Fried parameter estimation process, 
consider the images shown in Fig. \ref{boat_roi}.  These images show a region of interest for the Boats data with level 4 turbulence.  The truth image is shown in Fig. \ref{boat_roi}(a) and a single short exposure image is shown in Fig. \ref{boat_roi}(b).
The 300  frame average with no camera motion or registration is shown in Fig. \ref{boat_roi}(c).  Clearly this long exposure image
is much more blurred than the short exposure image.  The relationship between the long and short exposure images is
what allows us to estimate $r_0$.  When camera motion is present, 
global registration is required.   The average of the globally registered frames is shown in Fig. \ref{boat_roi}(d).
Note that there is less blurring in  Fig. \ref{boat_roi}(d) than in Fig. \ref{boat_roi}(c) due to the registration.  
This reduction in blurring is what leads to the inflated $r_0$ estimate, if left uncompensated.

\begin{figure}[t]
\begin{center}
{\includegraphics[trim=6cm 3cm 5cm 5cm,scale = .63]{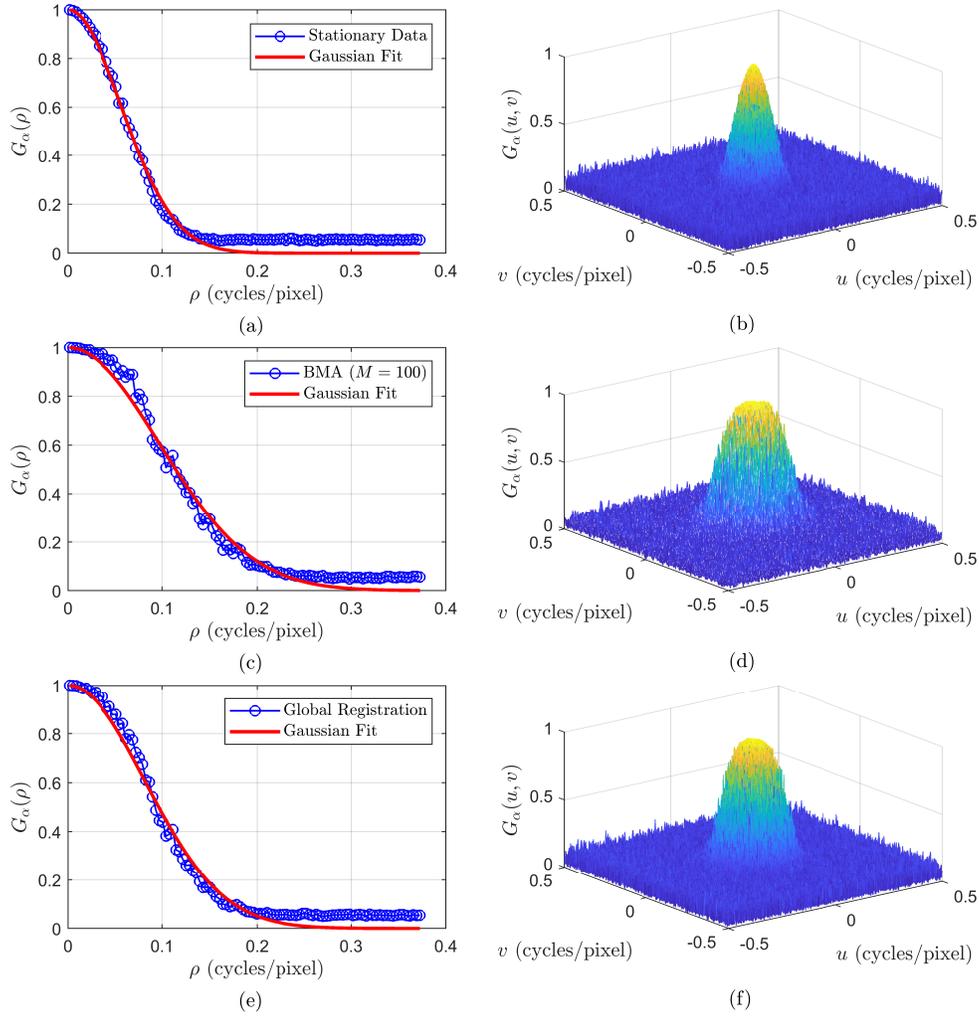}} 
\end{center}
\caption{Spectral ratio Gaussian fitting for level 4 turbulence defined in Table \ref{turbulence_parameters} for the image Boats.
Stationary camera data with no registration is shown in (a) and (b).  Data for BMA registered frames using $M=100$ are shown in (c) and (d).  Data for global registration are shown in (e) and (f).}
\label{gaussfit}
\end{figure}

Spectral ratios and the Gaussian fitting curves for the same data 
are shown in Fig. \ref{gaussfit} as a function of
spatial frequency in units of cycles per pixel spacing.
Figure \ref{gaussfit}(a) shows the spectral ratio radial function
 for the case of a stationary camera and no registration.  Here, the 
long exposure image has a low cut-off frequency, and the spectral ratio has a small $\sigma_G$.
In Fig. \ref{gaussfit}(c),  the spectral ratio is shown where BMA registration with 
$M=100$ is applied for camera motion compensation. 
The modified long exposure image is now less blurred, and this leads to a larger $\sigma_G$.
Finally in Fig. \ref{gaussfit}(e), we see the spectral ratio for global registration.  Here $\sigma_G$ is smaller
than with BMA, but still larger than with no registration.   By using the appropriate $\alpha$ for each scenario, all of these
can lead to effective $r_0$ estimates.

\subsubsection{Turbulence Mitigation with Simulated Data}

In this section we examine turbulence mitigation using the OTF model in Section \ref{otfappsec} with the $\alpha$ defined
in Section \ref{corr2dsec} and the spectral ratio $r_0$ estimate from Section \ref{ratio_section}.
Performance is measured in terms of peak signal-to-noise ratio (PSNR) and the structural similarity index (SSIM)~\cite{ssim2004}. In both cases, a larger number represents a better restoration.  In all cases the $r_0$ used for these restorations comes from the spectral ratio estimate assuming camera motion and global registration, as listed in Tables \ref{r0_error_table_boats} and \ref{r0_error_table_stream}.  The $\alpha$ employed is based on the type of registration.  For global registration results we use the average value of $\alpha=0.5252$ that corresponds to the full image size of $M=250$.   
For BMA, we use $M=10$ and a corresponding value of $\alpha=0.8878$.
This $\alpha$ is computed using $\varepsilon=1/12$.   The Wiener filter NSR used is $\Gamma=0.001$.

\begin{figure}[p]
\begin{center}
{\includegraphics[trim=5cm 9cm 5cm 9cm,scale=.75]{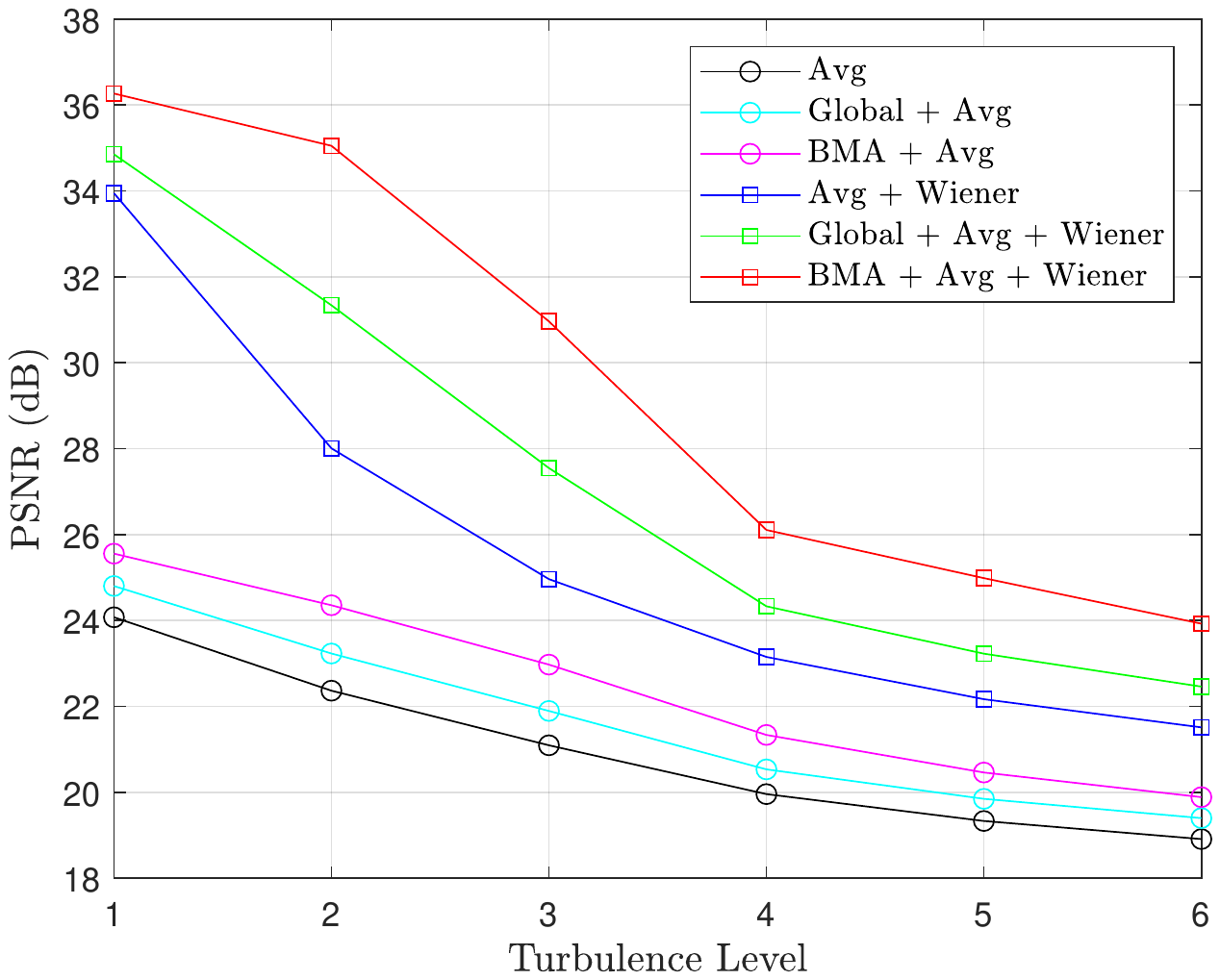}} 
\end{center}
\caption{Turbulence mitigation quantitative error analysis using the PSNR (dB) metric for the 6 turbulence levels in Table \ref{turbulence_parameters}  using the image Boats.
The $r_0$ value used is from the global registration spectral ratio estimate for this image sequence.}
\label{PSNRplot}

\begin{center}
{\includegraphics[trim=5cm 9cm 5cm 9cm,scale=.75]{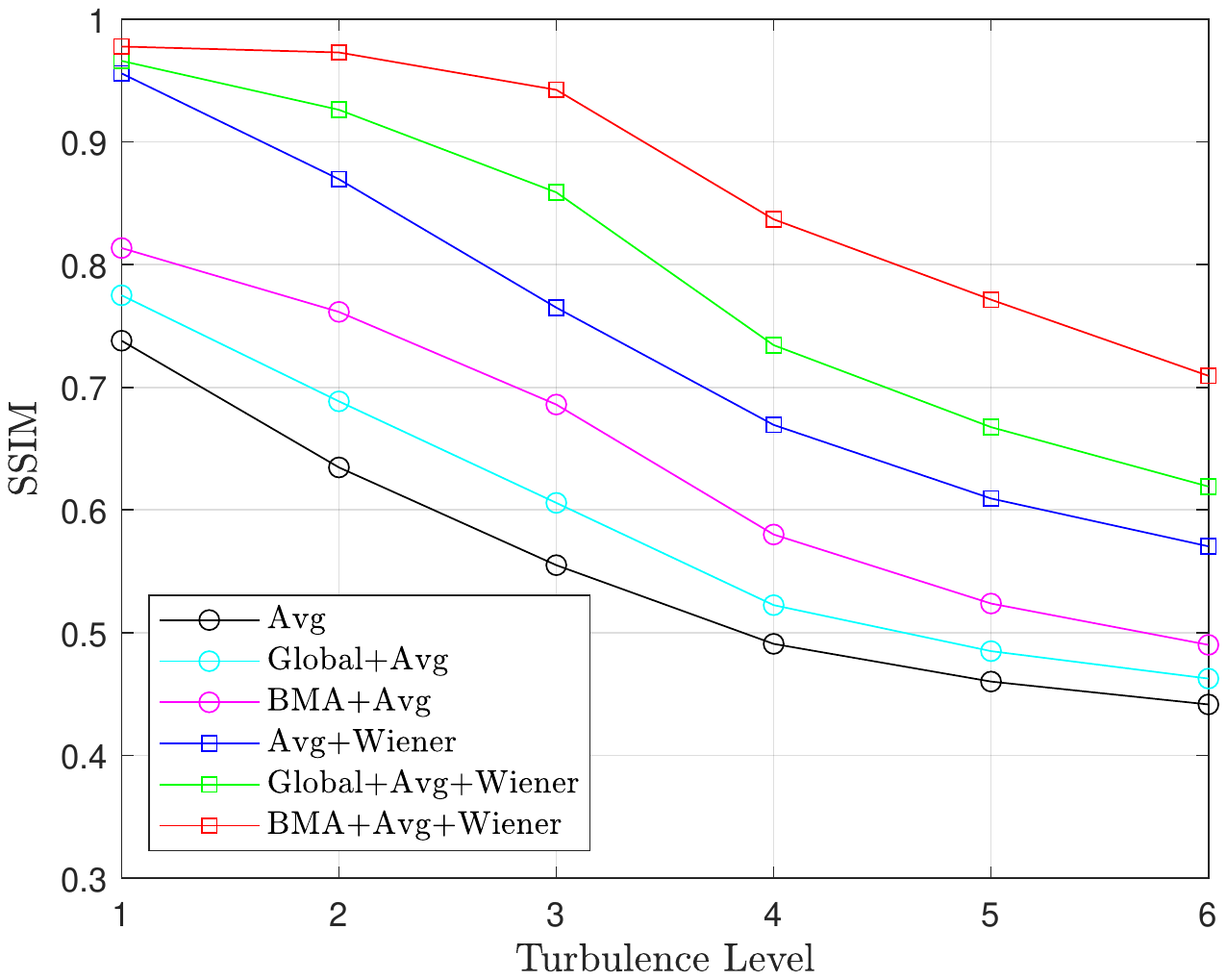}} 
\end{center}
\caption{Turbulence mitigation quantitative error analysis using the SSIM metric for the 6 turbulence levels in Table \ref{turbulence_parameters} using the image Boats.
The $r_0$ value used is from the global registration spectral ratio estimate for this image sequence.}
\label{SSIMplot}
\end{figure}

The PSNR results for the two truth images are provided in Tables \ref{PSNRtable} and \ref{PSNRtable2}, respectively.
The SSIM results are shown in Tables \ref{SSIMtable} and \ref{SSIMtable2}.   The PSNR and SSIM results for the Boats image data
are plotted in Fig. \ref{PSNRplot} and \ref{SSIMplot}, respectively.  These results show that employing registration prior to averaging and Wiener filtering boosts performance.  Here BMA with $M=10$ is better than global registration 
because the increased tilt correction produces an image with less blurring for the Wiener filter to restore.

\begin{table}[t]  
\small
\caption{Turbulence mitigation PSNR (dB) results for the Boats image using the optical parameters in Table \ref{optical_parameters} 
and the 6 turbulence levels in Table \ref{turbulence_parameters}. 
The $r_0$ value used is from the global registration spectral ratio estimate for this image sequence.}
\vspace{-.2in}
\label{PSNRtable}
\begin{center}
\begin{tabular}{ |l|c|c|c|c|c|c| }  
\hline
 {\bf Turbulence Mitigation} & \multicolumn{6}{c|}{{\bf Turbulence Degradation}} \\
\cline{2-7}
{\bf Method} & {\bf Level 1} & {\bf Level 2} & {\bf Level 3}  & {\bf Level 4} & {\bf Level 5} & {\bf Level 6}  \\ 
 \hline \hline 
Frame 1 & 22.8144 & 21.0166 & 19.5873  &  19.4271  & 19.3396 & 17.1386  \\ \hline  
Avg & 24.0733 & 22.3642 & 21.0926  &  19.9563  & 19.3304 & 18.9094  \\ \hline  
Global + Avg & 24.8010 & 23.2296 & 21.8923  &  20.5321  & 19.8463 & 19.3972  \\ \hline  
BMA ($M=10$) + Avg & 25.5550 & 24.3532 & 22.9716  &  21.3339  & 20.4592 & 19.8855  \\ \hline  
Avg + Wiener  & 33.9471 & 28.0064 & 24.9611  &  23.1497  & 22.1659 & 21.5099  \\ \hline  
Global + Avg + Wiener  & 34.8527 & 31.3373 & 27.5469  &  24.3288  & 23.2257 & 22.4591  \\ \hline  
BMA ($M=10$) + Avg + Wiener  & 36.2669 & 35.0499 & 30.9621  &  26.1062  & 24.9845 & 23.9241  \\ \hline  
 \end{tabular}
 \end{center}

\vspace{.2in}

 \small
 \caption{Turbulence mitigation PSNR (dB) results for the Stream and Bridge image using the optical parameters in Table \ref{optical_parameters} 
and the 6 turbulence levels in Table \ref{turbulence_parameters}.
The $r_0$ value used is from the global registration spectral ratio estimate for this image sequence.}
\vspace{-.1in}
\label{PSNRtable2}
\begin{center}
\begin{tabular}{ |l|c|c|c|c|c|c| }  
\hline
{\bf Turbulence Mitigation} & \multicolumn{6}{c|}{{\bf Turbulence Degradation}} \\
\cline{2-7}
{\bf Method} & {\bf Level 1} & {\bf Level 2} & {\bf Level 3}  & {\bf Level 4} & {\bf Level 5} & {\bf Level 6}  \\ 
 \hline \hline  
Frame 1 & 23.1948 & 22.0832 & 19.5952  &  20.0127  & 20.2346 & 17.5237  \\ \hline  
Avg & 24.1645 & 22.7428 & 21.6037  &  20.5968  & 20.0119 & 19.5737  \\ \hline  
Global + Avg & 24.7622 & 23.4439 & 22.2945  &  21.1216  & 20.4964 & 20.0397  \\ \hline  
BMA ($M=10$) + Avg & 25.4163 & 24.4391 & 23.2534  &  21.8289  & 21.0319 & 20.4842  \\ \hline  
Avg + Wiener  & 31.2960 & 27.0460 & 24.7814  &  23.1652  & 22.3452 & 21.8384  \\ \hline  
Global + Avg + Wiener  & 32.3660 & 29.2104 & 26.4898  &  24.0227  & 23.0100 & 22.3299  \\ \hline  
BMA ($M=10$) + Avg + Wiener  & 33.5384 & 32.3853 & 28.8911  &  25.5475  & 24.2221 & 23.1938  \\ \hline  
 \end{tabular}
 \end{center}
\end{table}

\begin{table}[t] 
\small
\caption{Turbulence mitigation SSIM results for the Boats image using the optical parameters in Table \ref{optical_parameters} 
and the 6 turbulence levels in Table \ref{turbulence_parameters}.
The $r_0$ value used is from the global registration spectral ratio estimate for this image sequence.}
\vspace{-.2in}
\label{SSIMtable}
\begin{center}
 \begin{tabular}{ |l|c|c|c|c|c|c| }  
\hline
{\bf Turbulence Mitigation}  & \multicolumn{6}{c|}{{\bf Turbulence Degradation}} \\
\cline{2-7}
{\bf Method} & {\bf Level 1} & {\bf Level 2} & {\bf Level 3}  & {\bf Level 4} & {\bf Level 5} & {\bf Level 6}  \\ 
 \hline \hline 
Frame 1 & 0.6965 & 0.6269 & 0.5259  &  0.5001  & 0.5003 & 0.4328  \\ \hline  
Avg & 0.7379 & 0.6347 & 0.5550  &  0.4909  & 0.4602 & 0.4415  \\ \hline  
Global + Avg & 0.7749 & 0.6885 & 0.6058  &  0.5225  & 0.4850 & 0.4626  \\ \hline  
BMA ($M=10$) + Avg & 0.8134 & 0.7614 & 0.6859  &  0.5800  & 0.5238 & 0.4901  \\ \hline  
Avg + Wiener  & 0.9557 & 0.8695 & 0.7649  &  0.6693  & 0.6094 & 0.5704  \\ \hline  
Global + Avg + Wiener  & 0.9659 & 0.9261 & 0.8588  &  0.7343  & 0.6675 & 0.6191  \\ \hline  
BMA ($M=10$) + Avg + Wiener  & 0.9776 & 0.9728 & 0.9424  &  0.8368  & 0.7714 & 0.7093  \\ \hline  
 \end{tabular}
 \end{center}

\vspace{.2in}

\small
\caption{Turbulence mitigation SSIM results for the Stream and Bridge image using the optical parameters in Table \ref{optical_parameters} 
and the 6 turbulence levels in Table \ref{turbulence_parameters}.
The $r_0$ value used is from the global registration spectral ratio estimate for this image sequence.}
\vspace{-.1in}
\label{SSIMtable2}
\begin{center}
\begin{tabular}{ |l|c|c|c|c|c|c| }  
\hline
{\bf Turbulence Mitigation} & \multicolumn{6}{c|}{{\bf Turbulence Degradation}} \\
\cline{2-7}
{\bf Method} & {\bf Level 1} & {\bf Level 2} & {\bf Level 3}  & {\bf Level 4} & {\bf Level 5} & {\bf Level 6}  \\ 
 \hline \hline 
Frame 1 & 0.6142 & 0.5320 & 0.4143  &  0.3962  & 0.3940 & 0.3323  \\ \hline  
Avg & 0.6470 & 0.5318 & 0.4452  &  0.3831  & 0.3547 & 0.3379  \\ \hline  
Global + Avg & 0.6896 & 0.5896 & 0.4983  &  0.4130  & 0.3778 & 0.3562  \\ \hline  
BMA ($M=10$) + Avg & 0.7373 & 0.6774 & 0.5916  &  0.4746  & 0.4151 & 0.3811  \\ \hline  
Avg + Wiener  & 0.9330 & 0.8214 & 0.6909  &  0.5811  & 0.5107 & 0.4734  \\ \hline  
Global + Avg + Wiener  & 0.9473 & 0.8918 & 0.8023  &  0.6460  & 0.5637 & 0.5120  \\ \hline  
BMA ($M=10$) + Avg + Wiener  & 0.9597 & 0.9536 & 0.9065  &  0.7768  & 0.6794 & 0.5889  \\ \hline 
 \end{tabular}
 \end{center}
\end{table}

\begin{figure}[t]
\begin{center}
{\includegraphics[trim=5cm 9cm 5cm 8cm,width=5.25in]{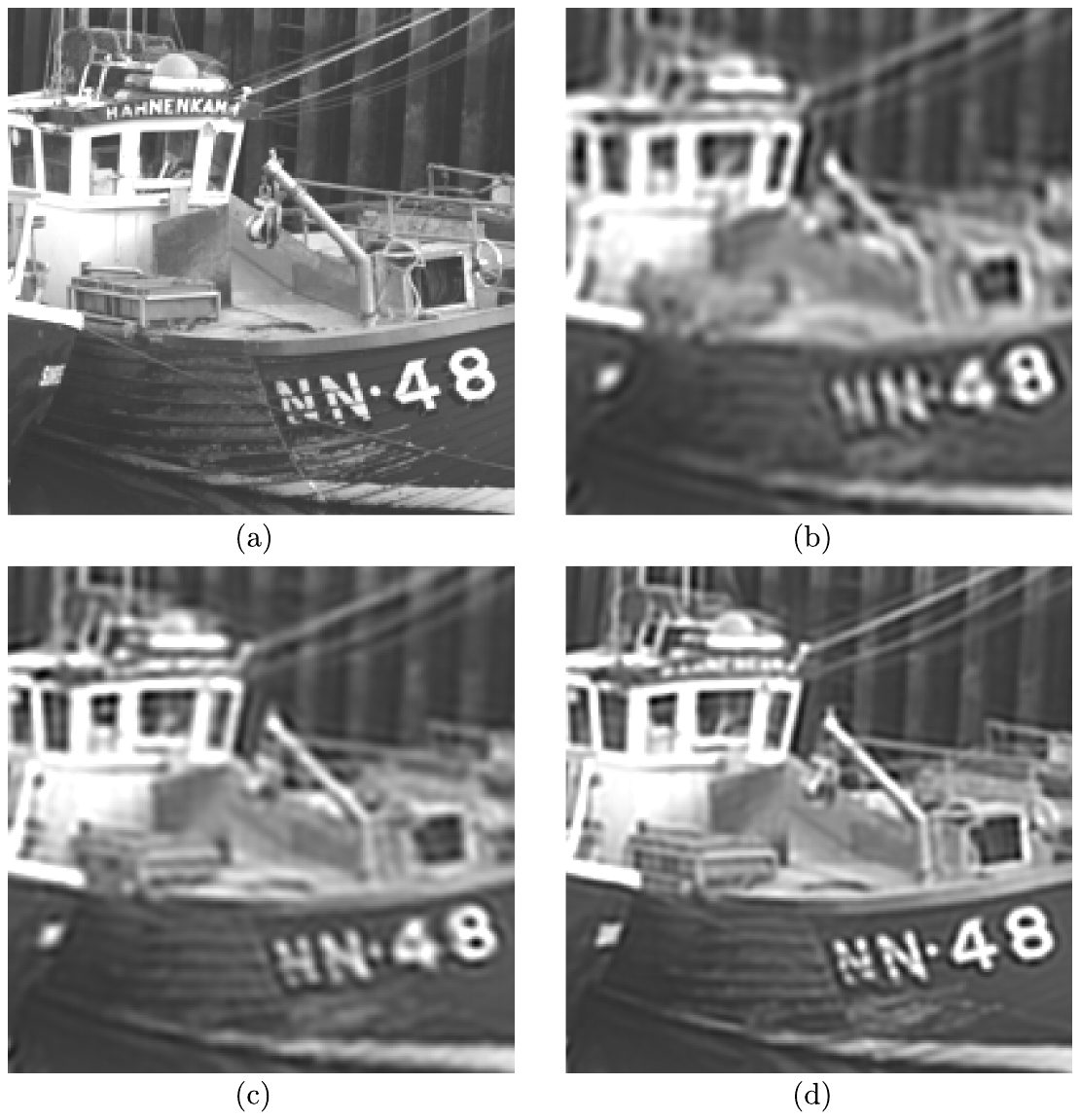}} 
\end{center}
\caption{Turbulence mitigation image results for level 4 turbulence from Table \ref{turbulence_parameters} with the image Boats.  
Regions of interest are shown for: (a) truth, (b) average + Wiener, (b) global registration + average + Wiener, (d) BMA ($M=10$) + average + Wiener.}
\label{boat_roi_tm}
\end{figure}

Image results for turbulence mitigation are shown in Fig. \ref{boat_roi_tm} for the Boats image data.
The truth region of interest is shown in Fig. \ref{boat_roi_tm}(a).  The average frame (i.e., long exposure) with Wiener filter applied is shown in Fig. \ref{boat_roi_tm}(b).  
The average of the globally registered frames with Wiener filter applied is shown in
Fig. \ref{boat_roi_tm}(c).   
Finally,  the average of $M=10$ BMA registered frames with Wiener filter applied is shown in 
Fig. \ref{boat_roi_tm}(d).  
Note that the best restored image detail is seen in Fig. \ref{boat_roi_tm}(d), 
followed by Figs. \ref{boat_roi_tm}(c) and (b).  
The more comprehensive the registration, the less blurred the input is to the Wiener filter, 
and the better the final result.

\subsection{Real Camera Data}
\label{realsec}

While we believe the simulated results are compelling, we believe it is also important to
provide results for real sensor data.
These results come from a camera with the specifications listed in Table \ref{optical_parameters2}.
An image sequence containing 600 frames is acquired from a stationary camera imaging 
a trailer and test pattern at a range of 1.8701 km.  
The image data are cropped to a size of $1001 \times 1001$ for subsequent processing.
An example image is shown in the results presented in Section \ref{realTM}.

\begin{table}[t]
\caption{Optical parameters for the real camera data.}
\vspace{-.2in}
\label{optical_parameters2}
\begin{center}
\begin{tabular}{ | l | l | } 
 \hline
 {\bf Parameter} & {\bf Value} \\
 \hline
Aperture & $D = 0.1333$ m  \\  
Focal length &  $l = 3.7320$ m  \\  
F-number &   $f/\# = 28$   \\
Wavelength &  $\lambda = 0.5500$ $\mu$m  \\
Object distance & $L=1870.1$  m \\ 
Actual pixel spacing (focal plane) & $\delta = 6.4500$  $\mu$m \\ 
 \hline
  \end{tabular}
   \end{center}
\end{table}

\subsubsection{Fried Parameter Estimation with Real Camera Data}

\begin{figure}[t]
\begin{center}
{\includegraphics[trim=2cm 10cm 1cm 10cm,width=5in]{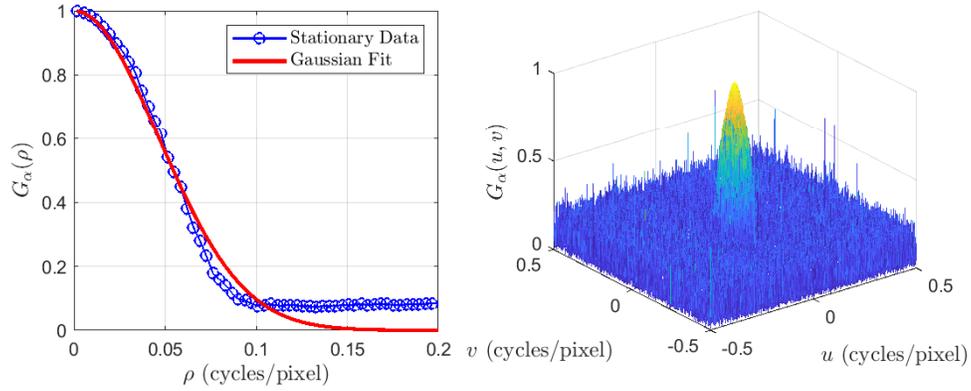}} 
\end{center}
\caption{Spectral ratio data for the 600 frames real camera image sequence.   Radial data with Gaussian curve fitting is shown on the left.  The 2D spectral ratio data is shown on the right. }
\label{gaussfit_mza}
\end{figure}

\begin{figure}[t]
\begin{center}
{\includegraphics[trim=5cm 9cm 5cm 9cm,scale=.75]{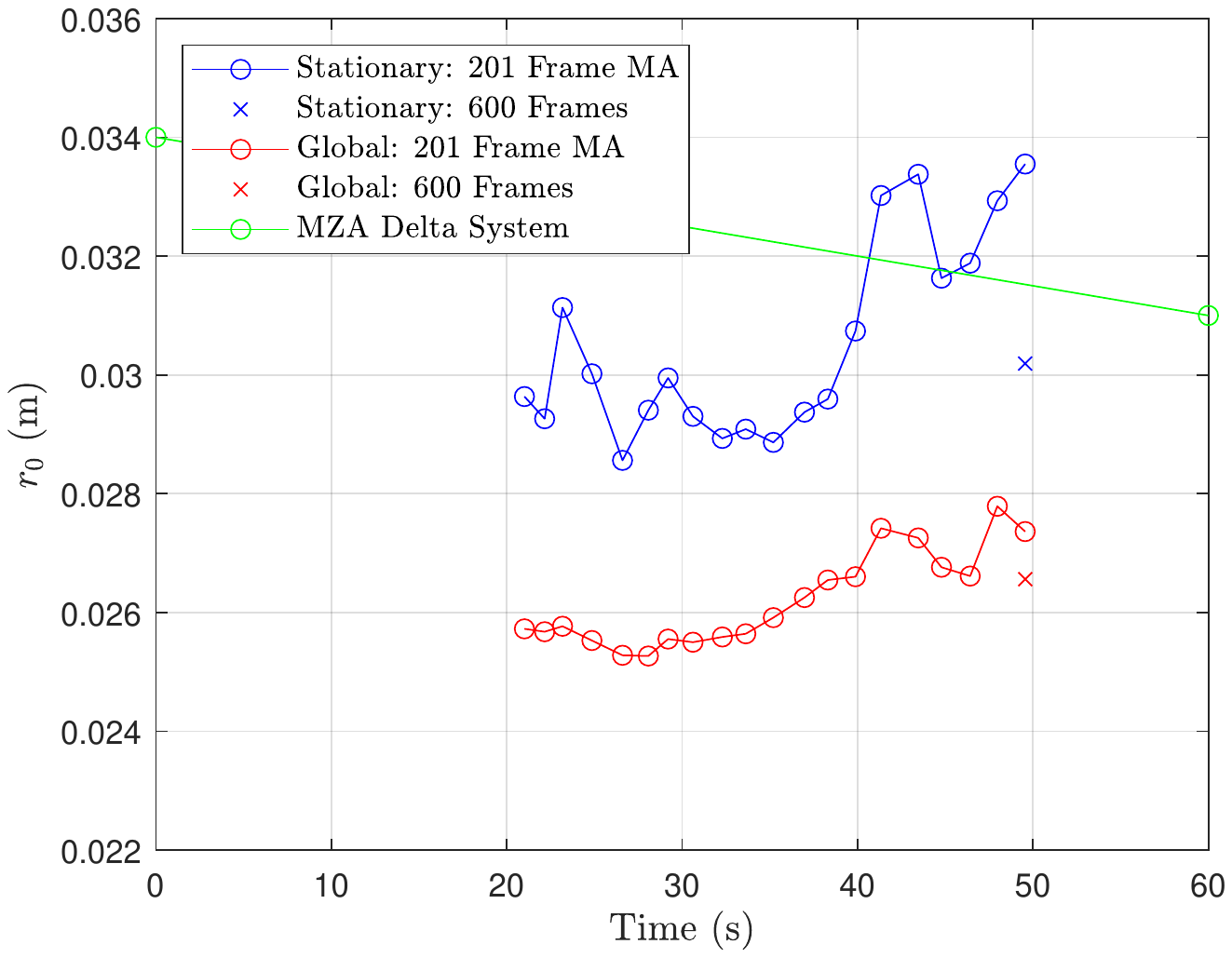}}  
\end{center}
\caption{Fried parameter estimation using the real camera image sequence.   ``Stationary'' results are with no camera motion and no registration.  ``Global'' results are using global image registration.  Ground truth is provided by the MZA Delta System.  }
\label{mza_r03}
\end{figure}

Our first experiment with the real data involves Fried parameter estimation.  The spectral ratio data for the stationary camera and no registration are shown in Fig. \ref{gaussfit_mza}.   These appear to be very similar to what we have observed with the simulated data.
From the spectral ratio and using $\alpha=0$, we obtain the ``Stationary'' $r_0$ estimates plotted in Fig. \ref{mza_r03}.  A single 
600 frame $r_0$ estimate is shown along with a temporal sequence of estimates using a 201 frame temporal moving window.
For reference, the $r_0$ measurement provided by the 
MZA Associates Delta System~\cite{mza_delta} is shown.
The nearest temporal MZA value for $r_0$ is $0.031$ m, and the Stationary spectral ratio estimate is  $0.0302$ m.  
This represents only a $-2.61$ \% difference, which is in keeping with the errors seen using the simulated data.

When global registration is applied to the acquired frames to emulate what would be required with camera motion, we obtain a different set of results.  For these results we compute the global registration average tilt correction factor for $M=500$ yielding $\alpha=0.5077$.  The corresponding $r_0$ estimates are shown in Fig. \ref{mza_r03} as ``Global'' estimates.  Both a 600 frame estimate and sequence of 201 frame moving window results are shown.   The 600 frame global registration estimate is $ 0.0266$ m.  This represents a $-14.32$\% difference from the MZA value.  We believe this increased error here may be largely the result of a highly variable $C_n^2(z)$ profile.  Estimated MZA profile data suggests that there is heavier turbulence close to the object.  
As shown in Fig. \ref{alpha_sensitivity_fig}, a constant path assumption leads to an inflated $\alpha$ in this scenario.  
In turn, this leads to an underestimate of $r_0$ based on Eq. (\ref{r0sigmaG}).
Even with this effect, the global registration based $r_0$ estimate may still be accurate enough for 
many applications.  Note that the stationary $r_0$ estimate is not adversely impacted by $C_n^2(z)$ profile variation in this way because no tilt correction factor is needed in that case.

\subsubsection{Turbulence Mitigation with Real Camera Data}\label{realTM}

\begin{figure}[p]
\begin{center}
{\includegraphics[trim=2cm 2cm 2cm 1cm,width=5.25in]{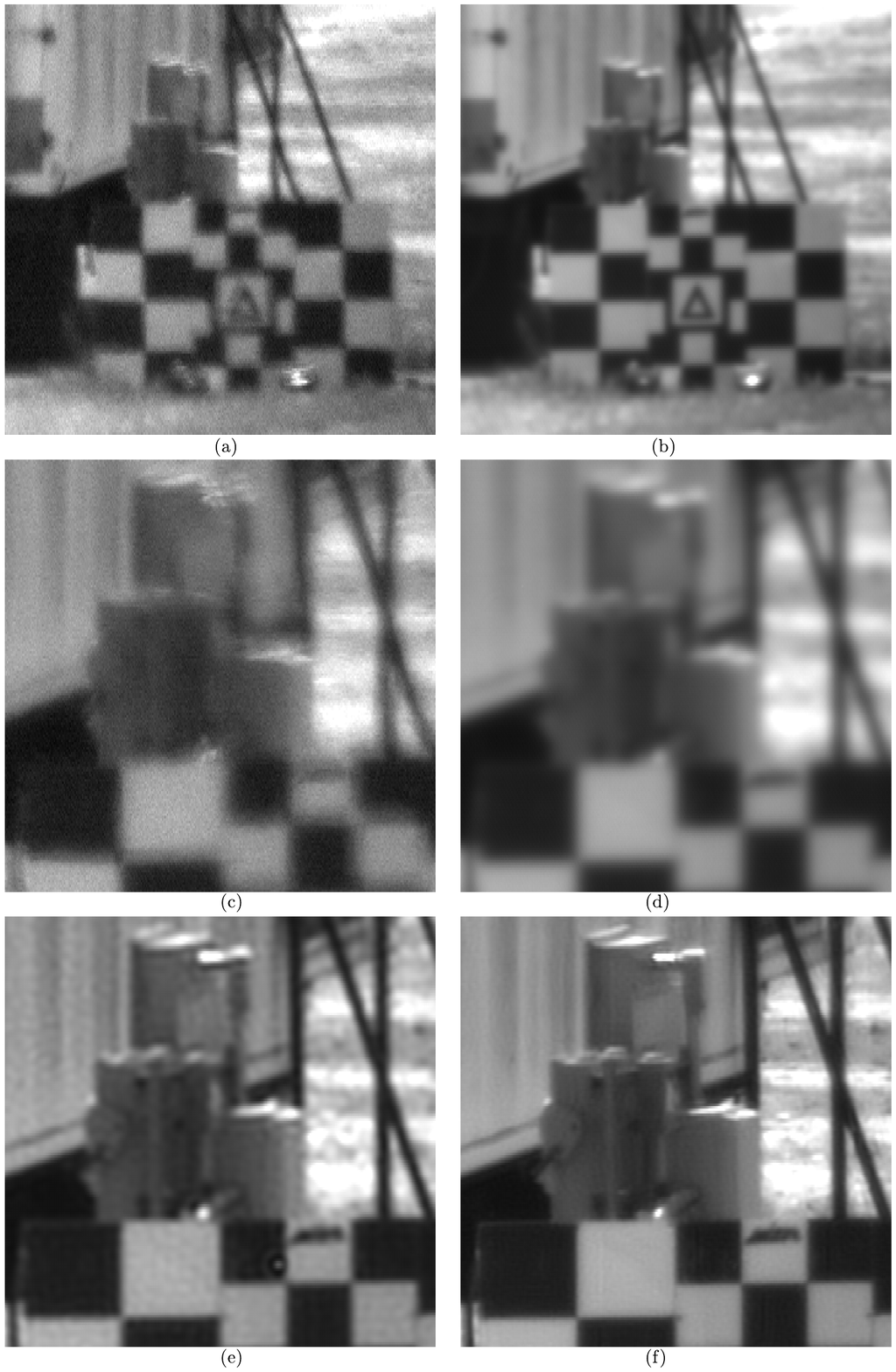}} 
\end{center}
\caption{Turbulence mitigation image results for the real camera data. (a) Short exposure Frame 1, (b) 600 frame average, (c) region of interest for short exposure Frame 1, (d)  region of interest for the 600 frame average, (e)  average + Wiener filter, (f) BMA ($M=15$) + average + Wiener filter.}
\label{delta_target_roi_tm}
\end{figure}

Turbulence mitigation image results using the real camera data are shown in Fig. \ref{delta_target_roi_tm}.
A single short exposure frame is shown in Fig. \ref{delta_target_roi_tm}(a).  The 600 frame temporal average is shown in
Fig. \ref{delta_target_roi_tm}(b).   Zoomed-in regions for these two images are shown in Figs. \ref{delta_target_roi_tm}(c) and (d), respectively.   The average frame with Wiener filter is shown in Fig. \ref{delta_target_roi_tm}(e).  
The output of the Wiener filter operating on the average of BMA ($M=15$)
registered frames is shown in Fig. \ref{delta_target_roi_tm}(f).  The stationary spectral ratio
estimate of $\hat{r}_0=0.0302$ m is used.  For the result in  Fig. \ref{delta_target_roi_tm}(e), we use $\alpha=0$.
For the result in Fig. \ref{delta_target_roi_tm}(f) we use the calculated value of $\alpha=0.8958$ that corresponds to 
$M=15$ and $\varepsilon=1/12$.  Note the improvement in Fig. \ref{delta_target_roi_tm}(f) using 
BMA registration over the result in Fig. \ref{delta_target_roi_tm}(e).  By including registration the restored image
appears sharper and appears to have less noise.   The extra noise reduction is from fixed pattern noise attenuation 
that results from averaging shifted frames~\cite{HardieNUC2000}.

\section{Conclusions}
\label{conclusions_section}

In this paper, we have presented turbulence tilt correlation statistics for two point sources as a function of the
parallel and perpendicular separation distances.  We converted these into 2D WSS autocorrelation functions characterizing the
random 2D tilt fields from turbulence.  We believe these help to inform a variety turbulence modeling and mitigation 
applications.  One way these statistics can be helpful is in understanding and modeling the impact of various image registration 
methods.  We are able to model the residual tilt variance in the case of patch based image registration such as BMA and global image registration.  We model these registration operations as LSI filters applied to the random tilt fields.  The output random process
has reduced tilt variance.  We define the tilt correction factor as one minus the ratio of the residual tilt variance to total input tilt variance.  We have shown that this tilt correction factor is only a function of the optical parameters and block size $M$ for a constant $C_n^2(z)$ path.  This gives us the ability to understand and quantify how effective image registration will be for different patch or image sizes defined by $M$, independent from the potentially unknown level of turbulence.

We use the tilt correction factor to model an atmospheric OTF that includes the impact of image registration.  This OTF model is then used to develop a modified spectral ratio Fried parameter estimation algorithm that can accommodate camera motion.   
Our method for Fried parameter estimation uses all of the pixels in the acquired images and does not require specialized sources or target images.   
With the ability to estimate the Fried parameter and determine the tilt correction factor, a complete OTF model can be defined and used for turbulence mitigation.    

The experimental results with simulated data show quantitatively that the Fried parameter estimation is very accurate over a wide range of turbulence levels on the datasets studied.  Results with real data in the stationary camera case are in good agreement with measured ground truth.  When registration is employed, higher errors were observed, and this is likely due to a non-constant $C_n^2(z)$ profile.
The goal of image registration here is to correct for camera motion with the least impact on the turbulence and the smallest tilt correction factor possible.  We have found that global registration can be used to compensate for camera motion and still preserve enough turbulence information to be able to effectively estimate $r_0$, provided the appropriate tilt correction parameter is employed.    

For BMWF turbulence mitigation, we have shown that our OTF model with estimated $r_0$ and tilt 
correction factor is highly effective.  We believe the quantitative results with simulated data, and the real camera data results, support this conclusion.  In this application, the goal of image registration is to perform the most atmospheric tilt correction possible.  
A high level of tilt correction produces a fused image with less turbulence motion blurring as the starting point for the Wiener filter restoration.  This is achieved by employing a BMA registration algorithm with a relatively small $M$.  When too small an $M$ is used, registration errors limit performance.  If too large an $M$ is used, less tilt correction is achieved.  Thus, a balance must be achieved between these two factors.

\section*{Funding}

This work has been supported in part under AFRL Award No. FA8650-17-D-1801.
Approved for public release under case number AFRL-2020-0563. 

\section*{Acknowledgments}

The authors would like to thank Dr. Barry Karch at AFRL for supporting this project and providing technical feedback.
Thanks to Joe French at Leidos for technical feedback and project management support.  
We also thank Bruce Wilcoxen, Amanda Caplinger, and Julie Tollefson with Leidos 
for project management support.  
Thank you to the engineering teams at Leios and MZA Associates for their roles in acquiring the real data used here.
Thanks to Yakov Diskin at MZA Associates for kindly providing information related to 
their Delta system for atmospheric characterization.

\section*{Disclosures}

The authors declare no conflicts of interest.

\section*{Data availability} 
The truth images used here may be obtained in~\cite{Agustsson_2017_CVPR_Workshops,usc_images}. 
The data with turbulence were simulated using the method in~\cite{HardieSimulation2017}.
The real camera data are not publicly available at this time.



\end{document}